\documentclass[journal]{IEEEtran}
\usepackage{graphicx,amssymb,lineno}
\usepackage{amsmath,amsfonts,amssymb}
\usepackage{cases}
\usepackage{balance}

\usepackage{algorithm}
\usepackage{algorithmic}
\usepackage[usenames]{color}
\usepackage{subfigure}
\usepackage{setspace}
\usepackage{bigstrut}
\usepackage{multirow}
\usepackage{array}
\usepackage{booktabs}
\usepackage{balance}
\begin{document}

\title{Hybrid Satellite-Terrestrial Communication Networks for the Maritime Internet of Things: \\Key Technologies, Opportunities, and Challenges}
%}
\author{Te Wei, Wei Feng, \emph{Senior Member, IEEE}, Yunfei Chen, \emph{Senior Member, IEEE},\\ Cheng-Xiang Wang, \emph{Fellow, IEEE}, Ning Ge, \emph{Member, IEEE}, and Jianhua Lu, \emph{Fellow, IEEE}
\thanks{
This work was supported in part by the National Natural Science Foundation of China (Grant No. 61922049, 61771286, 61941104, 61960206006, 61701457, 91638205), the National Key R\&D Program of China under Grant 2018YFA0701601, the Frontiers Science Center for Mobile Information Communication and Security, the High Level Innovation and Entrepreneurial Research Team Program in Jiangsu, the High Level Innovation and Entrepreneurial Talent Introduction Program in Jiangsu, the Research Fund of National Mobile Communications Research Laboratory, Southeast University, under Grant 2020B01, the Fundamental Research Funds for the Central Universities under Grant 2242020R30001, the Huawei Cooperation Project, the EU H2020 RISE TESTBED2 project under Grant 872172, the Nantong Technology Program under Grant JC2019115, and the Beijing Innovation Center for Future Chip.

T. Wei, W. Feng (corresponding author), N. Ge, and J. Lu are with the Beijing National Research Center for Information Science and Technology, Department of Electronic Engineering, Tsinghua University, Beijing 100084, China. T. Wei is also with the Department of WLAN Development, Huawei Beijing Research Center, Beijing 100085, China
%. W. Feng is also with the Peng Cheng Laboratory, Shenzhen 518000, China
(e-mail: wei-t14@tsinghua.org.cn;~fengwei@tsinghua.edu.cn;~gening@tsinghua.edu.cn;~lhh-dee@tsinghua.edu.cn).

Y. Chen is with the School of Engineering, University of Warwick, Coventry CV4 7AL, U.K. (e-mail: Yunfei.Chen@warwick.ac.uk).

%C.-X. Wang is with the National Mobile Communications Research Laboratory, School of Information Science and Engineering, Southeast University, Nanjing 210096, P. R. China (e-mail: chxwang@seu.edu.cn).

C.-X. Wang is with the National Mobile Communications Research Laboratory, School of Information Science and Engineering, Southeast University, Nanjing, 210096, China, and also with the Purple Mountain Laboratories, Nanjing 211111, China (e-mail: chxwang@seu.edu.cn).

%C.-X. Wang is with the National Mobile Communications Research Laboratory, School of Information Science and Engineering, Southeast University, Nanjing 210096, China. He is also with the Purple Mountain Laboratories, Nanjing 211111, China (e-mail: chxwang@seu.edu.cn).

%J. Wang is with the School of Electronic and Information Engineering, Nantong University, Nantong 226019, P. R. China.~(e-mail: wangjue@ntu.edu.cn).
}
}

%\markboth{IEEE INTERNET OF THINGS JOURNAL, VOL. XX, NO. XX, MONTH 2020}%

\maketitle

%\begin{spacing}{1.51}
\begin{abstract}
With the rapid development of marine activities, there has been an increasing number of Internet of Things (IoT) devices on the ocean. This leads to a growing demand for high-speed and ultra-reliable maritime communications. It has been reported that a large performance loss is often inevitable if the existing fourth-generation (4G), fifth-generation (5G) or satellite communication technologies are used directly on the ocean. Hence, conventional theories and methods need to be tailored to this maritime scenario to match its unique characteristics, such as dynamic electromagnetic propagation environments, geometrically limited available base station (BS) sites and rigorous service demands from mission-critical applications. Towards this end, we provide a survey on the demand for maritime communications enabled by state-of-the-art hybrid satellite-terrestrial maritime communication networks (MCNs). We categorize the enabling technologies into three types based on their aims: enhancing transmission efficiency, extending network coverage, and provisioning maritime-specific services. Future developments and open issues are also discussed. Based on this discussion, we envision the use of external auxiliary information, such as sea state and atmosphere conditions, to build up an environment-aware, service-driven, and integrated satellite-air-ground MCN.

\end{abstract}

\begin{IEEEkeywords}
Maritime communication network, maritime channel, maritime service, satellite-air-ground integration, knowledge library.
\end{IEEEkeywords}
%\end{spacing}

%\begin{spacing}{1.01}
%\begin{spacing}{1.9}
\section{Introduction}\label{S1}

%application requirements

Maritime activities, such as marine tourism, offshore aquaculture, and oceanic mineral exploration, have seen rapid development in recent years. With the increasing number of vessels, offshore platforms, buoys, etc., there has been a growing demand for high-speed and ultra-reliable maritime communications to connect them \cite{p212}--\cite{p21207252}.
%With the rapid development of marine activities, such as marine tourism, offshore aquaculture, and oceanic mineral exploration, there has been a growing demand for high-speed and ultra-reliable maritime communication services \cite{p212}.
%Oceanic economic and cultural exchanges between countries, such as the Maritime Silk Road project in China, have further promoted this demand \cite{p102}.
%China has a vast sea area with more than 10,000 ships sailing daily, and its offshore economic activity is very active. In 2016, China's total marine economic output exceeded 7 trillion yuan, accounting for approximately 9.5\% of China's GDP[1]. The construction of the Maritime Silk Road in China has also further promoted the development of the marine economy and promoted the economic and cultural exchanges between the coastal countries [2].
%At the same time, with the rapid development of marine activities such as marine tourism, offshore aquaculture, and marine mineral development, the demand for high-speed and reliable maritime communication services is continuously increasing.
For example, navigational information and operational data are required for the safe navigation of all vessels, and multi-media communication services are needed for passengers, crew, and fishermen onboard.
Similarly, offshore drilling platforms require real-time operational data communications, and buoys also have a large amount of meteorological and hydrological data to upload \cite{p1802}--\cite{p1803}.
For maritime rescue, in addition to information exchange using texts and voices, real-time videos are often required for better ship-to-ship and ship-to-shore coordination \cite{p1805}.
Therefore, building a broadband maritime communication network (MCN) for the maritime Internet of Things (IoT) is of great significance for marine transportation \cite{p104}\cite{p1502}, production safety \cite{p1504} and emergency rescue \cite{p1505}.

\begin{table*}[t]
	\centering
	\caption{Comparison between cellular and maritime communications.}
	\label{tab:table01}%
	\scriptsize
	\begin{tabular}{|c|p{22em}|p{30em}|}
		\hline
\multicolumn{1}{|c|}{\textbf{Characteristics}} & \textbf{Cellular communications} & \textbf{Maritime communications} \bigstrut\\
		\hline
		\multicolumn{1}{|c|}{\multirow{3}[4]{*}{Single BS coverage}} & Small  & Wide \bigstrut[t]\\
            & 4G: 500--2000 m for a single cell in urban area & Shore-based MCN: 10--100 km for a single BS \\
            & 5G: 100--300 m for a single cell in urban area & MCN using ship-borne/UAV-enabled BSs: 1--50 km for a single BS \\		
            & (Achieved by building a large number of BSs) &  (Due to the limited number of geographically available BSs) \bigstrut[b]\\		
		\hline		
		\multicolumn{1}{|c|}{\multirow{3}[18]{*}{Wireless channel}} & Lower propagation loss\newline{} (Due to small cell radius\bigstrut) & Higher propagation loss\newline{}(Due to long-distance transmission \bigstrut)\\
		\cline{2-3}          & Mainly affected by blocks and scatteres \bigstrut& Mainly affected by sea surface conditions (such as tidal waves), and atmospheric conditions (such as temperature, humidity, and wind speed) \bigstrut\\
		\cline{2-3}          & Mostly multi-path channels \newline{}(Rician channels in open areas) & Mostly Rician channels (with a direct path, a specularly reflected path, several diffusely reflected paths, and several atmospheric scattering paths) \bigstrut\\
		\hline
		\multicolumn{1}{|c|}{\multirow{3}[6]{*}{One-way transmission delay}} & Low \newline{} 4G: less than 10 ms\newline{}5G: less than 1 ms & High\newline{} Onshore BSs: Similar to 4G/5G \newline{} GEO: approx. 270 ms\newline{}MEO: approx. 130 ms (e.g., for O3b)\newline{}LEO: less than 40 ms (e.g.,  10--30 ms for Globalstar) \bigstrut\\
		\hline
		\multicolumn{1}{|c|}{\multirow{3}[2]{*}{Service}} & Mainly for mobile communications services & Mainly for maritime affairs, fisheries, ports, shipping, and coastal defence \newline{} E.g., \textcolor{black}{accurate and intelligent navigational communications for all vessels, real-time operational data communications for offshore drilling platforms, high-throughput multimedia downloading services for passenger/crew infotainment, and emergency communications with low-latency and high-reliability for maritime rescue}\bigstrut\\
		\hline
	\end{tabular}%
\end{table*}%

%existing systems/projects &problems
Currently, mobile terminals on the ocean mainly rely on maritime satellites or base stations (BSs) on the coast/island to acquire services.
%Existing maritime communication means mainly include satellite communication, shore-based communication and ship-borne communication.
%In terms of satellite communications:
Narrow-band satellites, represented by International Maritime Satellites (Inmarsat), mainly provide services such as telephone, telegraph, and fax, at a low communication rate. For example, the annual throughput of Inmarsat was only 66 Gbps in 2016, while the number of ships has exceeded 2 million. Thus, the average communication rate per ship is less than 33 kbps \cite{p1308}. To meet the demand for broadband satellite communication services, several companies have launched high-throughput satellites, such as EchoStar-19 (also known as Jupiter-2) by EchoStar and the Starlink project by SpaceX. %However, satellite communications are not cost-effective, due to the limited spectrum and orbit resources, and high operating costs. Therefore, satellite communication systems are not suitable for providing offshore broadband communication services.
%In terms of shore-based communication:
In addition to maritime satellites, shore \& island-based BSs are also used to extend the coverage of terrestrial fourth-generation (4G)/fifth-generation (5G) networks for maritime activities \cite{p2007}.
The existing shore-based communication systems, such as the Navigation Telex (NAVTEX) system and the Automatic Identification System (AIS), mainly provide services for information broadcasting, voice, and ship identification, but they cannot provide high-speed data services \cite{p202}.
To improve the communication rate, several companies, such as Huawei and Ericsson, have carried out long-distance shore-to-ship transmission tests based on Worldwide Interoperability for Microwave Access (WiMAX) or Long Term Evolution (LTE) networks \cite{p211}\cite{p1801}.
%Singapore has launched a WiMAX-based WISEPORT program. Huawei and Ericsson are also based on LTE networks. Long-distance transmission tests have been carried out on the coast near Qingdao, China.
%At the coastal elevated base stations, directional beams are generated through array antennas for fixed-point coverage, but does not yet support large-scale mobile communications.
However, the coverage of these systems is limited by the earth curvature and maritime environment.

To provide a practically affordable solution for broadband maritime communications, an efficient hybrid satellite-terrestrial MCN is urgently required to combine the advantage of satellites' wide coverage with shore-based systems' high capacity. It is believed that an arm-hand-like network architecture is beneficial to cover the widely but sparsely distributed maritime mobile terminals. In this framework, satellites and shore-based systems provide backhauls for dynamic global coverage (like the arms), while ship-to-ship interconnections and unmanned aerial vehicles (UAVs) can be exploited for enhancing local coverage (like the hands) \cite{p1507}.
%To enhance offshore coverage,
%Singapore has planned to build a mesh network through ship-to-ship interconnections, and unmanned aerial vehicles (UAVs) can also be exploited to serve as aerial maritime BSs \cite{p1507}.
%Despite all these approaches, there are still open issues towards the establishment of an efficient hybrid satellite-terrestrial MCN.
However, different from terrestrial cellular networks, the MCN still faces many challenges due to the complicated electromagnetic propagation environment, network topology patterns, and service demands from mission-critical applications \cite{p304}--\cite{p1028}:

$\cdot$ \emph{Transmission efficiency:}
Compared with the terrestrial environment, the atmosphere over the sea surface is unevenly distributed due to the large amount of seawater evaporation. Shore-to-ship and ship-to-ship communications are very vulnerable to both sea surface conditions, such as tidal waves, and atmospheric conditions, such as temperature, humidity, and wind speed.
In addition, the height and the angle of ship-borne antennas vary greatly with the ocean waves. Thus, the fading channel is particularly sensitive to parameters, such as antenna height and angle, which may cause frequent link interruption.
Therefore, the transmission efficiency in these applications is often low, due to these complicated time-varying factors.

$\cdot$ \emph{Coverage performance:}
%Satellite communications are less susceptible to maritime environment than shore/ship-to-ship communications.
%The integration of satellite and terrestrial systems can improve communication reliability and expand the coverage of MCNs.
In a terrestrial network, it is possible to increase the broadband coverage by installing more BSs.
However, in an MCN, the available BS sites are very limited.
Due to the limited onshore BS sites and the strong mobility of the ship-borne BSs, aerial BSs, and low-earth orbit (LEO) satellites, the topology of the hybrid satellite-terrestrial MCN is highly dynamic and irregular.
Blind zones always exist in the planned coverage area.
Additionally, if the BS covers remote users using high power, it will generate strong co-channel interference to the users served by neighbouring BSs.
%As the spectrum resources available for maritime communications are limited, the co-channel interference is inevitable.
%The existing of severely-interfered areas is also a hard nut to crack that restricts the broadband coverage of the MCN.
The coverage performance of MCNs is thus restricted by blind zones and areas with severe interference.

%It is difficult to build BSs on the sea except for deploying network nodes in fixed locations such as existing lighthouses. Besides, for onshore BSs, in order to achieve wide coverage, the antennas must be sufficiently high, so the BSs usually need to be set up on the high mountains along the coast, making the available BS sites even limited. In addition, the spectrum and orbital resources for maritime satellite communications are also very limited.
%%On the one hand, the coverage of a single station is large, and the transmission power of the base station usually needs to be set high and the energy efficiency is low.
%Therefore, there is a sharp contradiction between the coverage and the throughput of the MCN, and there always exist blind zones within the coverage.

$\cdot$ \emph{Service provisioning:}
%Besides,
Marine information network contains several industries, such as maritime affairs, fisheries, ports, shipping, and
coastal defence. Their maritime application scenarios are also quite different with unique service requirements.
%Currently, the fixed full coverage scheme is difficult to support all service requirements of broadband services.
Providing reliable services for all of these maritime-specific applications is a major challenge for the MCN.
%It is also important to effectively allocate spectrum and power resources based on the characteristics of different service requirements.
%is also a problem that needs studying.

\begin{figure*} [htb]
\begin{center}
\includegraphics*[width=18cm]{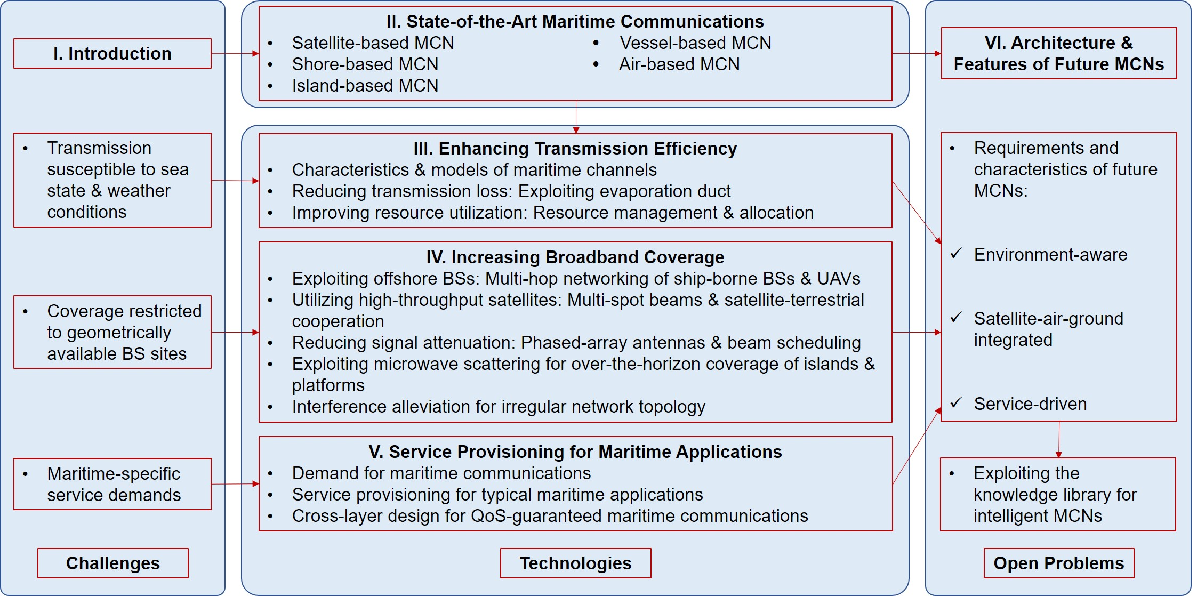} %这行为导入图片文件名称
\end{center}
%\vspace*{-4mm}
\caption{The structure of this paper.}\label{fig:2}  %这行为文章中显示图片的标题
%\vspace*{-2mm}
\end{figure*}

In Table I, we illustrate the difference between traditional cellular communications and relevant maritime communications.
%, which bring about the above-mentioned challenges faced by the MCN.
To address the unique challenges in maritime communications, conventional communications and networking theories and methods need to be tailored
for maritime scenarios, leading to an emerging area of communications.
%Maritime communications is emerging as an important research field, and
To date, a number of studies have been conducted on MCNs.
To enhance the maritime transmission efficiency, various channel measurement and modelling projects have been performed to analyse the impact of important system parameters (frequency, antenna height, etc.) and maritime environments (sea state, weather conditions, etc.) on the maritime channel. Moreover, advanced resource allocation schemes, such as dynamical beamforming and user scheduling techniques, have been studied to adapt to the dynamic changes in maritime channels. In addition, several studies have exploited the evaporation duct effect to improve the transmission efficiency, especially for remote ship-to-ship/shore transmissions.
To expand the network coverage, various BSs have been utilized, including onshore BSs, ship-borne BSs, aerial BSs, and satellites.
For these BSs, advanced beamforming and microwave scattering techniques have been studied to reduce the signal attenuation and extend the coverage. In addition, interference mitigation and satellite-terrestrial coordination schemes have been studied to overcome the interference due to the irregular network topology.
To satisfy the unique maritime service requirements, different systems and their transmission and resource allocation techniques have been studied for different service requirements, such as bandwidth, latency, and criticality.

Although there have been a large number of works on the above topics, there are very few survey papers on MCNs.
Additionally, most of them are focused on a specific issue, such as
channel models \cite{p304}--\cite{p10202}, network management \cite{p401}, or existing systems \cite{p221}--\cite{p1028}.
These issues are closely related to the characteristics of MCNs, but they were addressed separately without considering their interplays. For example, the surveys on maritime channel models have pointed out the challenges faced by environment-sensitive maritime channels but have not discussed any technologies to enhance transmission efficiency in such maritime scenarios.
Moreover, many other important issues for MCNs, such as resource allocation, service provisioning and network integration, are not completely discussed by any of these surveys.
Although the survey papers on some relevant topics, such as 5G channel measurements and models \cite{p3007}, space-air-ground integrated networks \cite{p3001}, \textcolor{black}{and cognitive-radio-based IoT \cite{p20201206-0}}, could shed light on the development of an efficient hybrid satellite-terrestrial MCN, in general they have not specialized in maritime scenarios.
To the best of the authors' knowledge, a survey paper dedicated to hybrid satellite-terrestrial MCNs with a complete picture of maritime communications
%covering various and the latest technologies and presenting the opportunities and challenges
is not available in the literature but is crucial to pave the way for
the understanding of the unique features of MCNs.
To fill in the gap, this paper provides a survey on maritime communications, which not only includes the topics that have not been previously covered, such as maritime service provisioning, but also extends the existing surveys by addressing the unique features of MCNs and the inner connections among the topics.

This paper provides a survey on the demand, state of the art, major challenges and key technical approaches in maritime communications. In particular, it focuses on the unique characteristics of maritime communications that are not seen in terrestrial or satellite communications. \textcolor{black}{It discusses the major challenges of MCNs due to unique meteorological conditions and geographical environments, as well as heterogeneous service requirements. Consequently, it illustrates the corresponding solutions from link-level, network-level, and service-level perspectives.}
Finally, it makes recommendations on developing an environment-aware, service-driven and satellite-air-ground integrated MCN, which is smart enough to utilize the external auxiliary information, e.g., the sea state conditions. The relevant open issues are also pointed out.

The remainder of this paper is organized as follows.
Section II briefly reviews the state-of-the-art MCNs, including satellite-based, shore-based, island-based, vessel-based, air-based, and underwater MCNs. In Section III, we introduce the key technologies to enhance maritime transmission efficiency.
%, including the characteristics and models of maritime channels, the use of evaporation duct, and efficient resource allocation.
Section IV introduces the key technologies to extend the coverage of MCNs.
%, such as multi-hop networking of ship-borne and UAV-enabled BSs, utilizing high-throughput satellites and satellite-terrestrial cooperation, dynamic beam scheduling, microwave scattering, and interference alleviation.
The demand for maritime communications, and key technologies for providing maritime-specific services such as low-power communications and cross-layer design, are discussed in Section V.
In Section VI, we suggest the architecture and features of future smart MCNs, as well as suggesting future research topics. Section VII concludes this paper. Figure 1 shows the outline of the paper.

%the research status of MCNs, and focuses on the unique characteristics of maritime communications compared with traditional terrestrial or satellite communications.
%This paper will uncover the complete picture of maritime communications, and pave the way for related researchers to understand the unique features of MCNs.

%This paper will provide maritime communication researchers and engineers with a comprehensive introduction to the demands, challenges, developing history and status quo, key technologies, and the insights of future developing routes of maritime communications, promoting the construction and development of SMCNs.

%\begin{table}
%\centering
%\caption{Maritime Communications}
%\scriptsize
%\begin{tabular}{|c|c|}  %通过添加|来表示是否需要绘制竖线
%\hline  % 在表格最上方绘制横线
%Service requirements & secure communication, proprietary communication and public communication\\
%\hline % 在表格最下方绘制横线
%Systems and projects & shore-based, sea-based, air-based and satellite-based systems and projects\\
%\hline % 在表格最下方绘制横线
%Enhancing maritime transmission efficiency & exploiting large-scale CSIT, microwave scattering and evaporation duct\\
%\hline % 在表格最下方绘制横线
%Ensuring wide-area coverage  & antennas and beamforming, multi-hop networking and high frequency transmission techniques\\
%\hline % 在表格最下方绘制横线
%QoS-guaranteed techniques & \\
%\hline % 在表格最下方绘制横线
%\end{tabular}
%\end{table}

%In order to meet the increasing demand, several maritime communication network (MCN) projects have been developed in recent years, e.g., the BLUECOM+ project, the MarCom project, and the TRITON project \cite{p321}--\cite{p32}.

\section{State-of-the-Art Maritime Communications}

Maritime communications began at the turn of the 20th century, pioneered by Marconi's work on long-distance radio transmissions.
In 1897, Marconi established a 6-km communication link across the Bristol Channel, which is the first wireless communication over open sea.
In 1899, he initiated the transmission across the English channel, from Wimereux, France to Dover, England, approximately 50 km away.
In the same year, Marconi and his assistants installed wireless equipment on the Saint Paul, a trans-Atlantic passenger liner, and successfully received telegrams from the coast station 122 km away.
In 1901, Marconi achieved trans-Atlantic communications with a transmission distance of over 3000 km, using a 20 kW high-power transmitter and a receiving antenna with a height of 150 metres \cite{p1051}--\cite{p1053}.
Marconi's experiments aroused great interest from the shipping industry in Europe and North America.
From then on, many countries began to install coast stations and ship-borne radio stations.
Narrowband communication services such as telegraph, telephone and fax were provided using data transmissions via intermediate frequency (MF, 0.3--3 MHz), high frequency (HF, 3--30 MHz), very high frequency (VHF, 30--300 MHz), and ultra high frequency (UHF, 0.3--3 GHz) bands. Among them, VHF is mostly used for radio and television broadcast. It is also a licensed band for aviation and navigation, which is important for the safe navigation of ships within 25 nautical miles along the coast.
VHF terminals have been widely used on merchant ships, fishing boats, official ships, yachts, and lifeboats. It is the most popular communication equipment for marine vessels \cite{p1019}--\cite{p6013}.

At present, several works have been conducted on broadband MCNs. Norway and Portugal launched the MARCOM project and the BLUECOM+ project, respectively, to provide broadband communications for remote areas by using Wi-Fi, General Packet Radio Service (GPRS), Universal Mobile Telecommunications System (UMTS), LTE technologies or their combination \cite{p201}\cite{p2003}. Singapore launched the TRITON project, where a wireless multi-hop network is formed between adjacent vessels, maritime beacons, and buoys, to ensure wide-area coverage \cite{p204}. In addition, the authors in \cite{p402}--\cite{p2004} discussed methods to achieve maritime communications through collaborative heterogeneous wireless networks using terrestrial networks, satellite networks, and other types of wireless networks. %Some other projects or solutions are listed as follows.

\begin{figure*} [t]
\begin{center}
\includegraphics*[width=13cm]{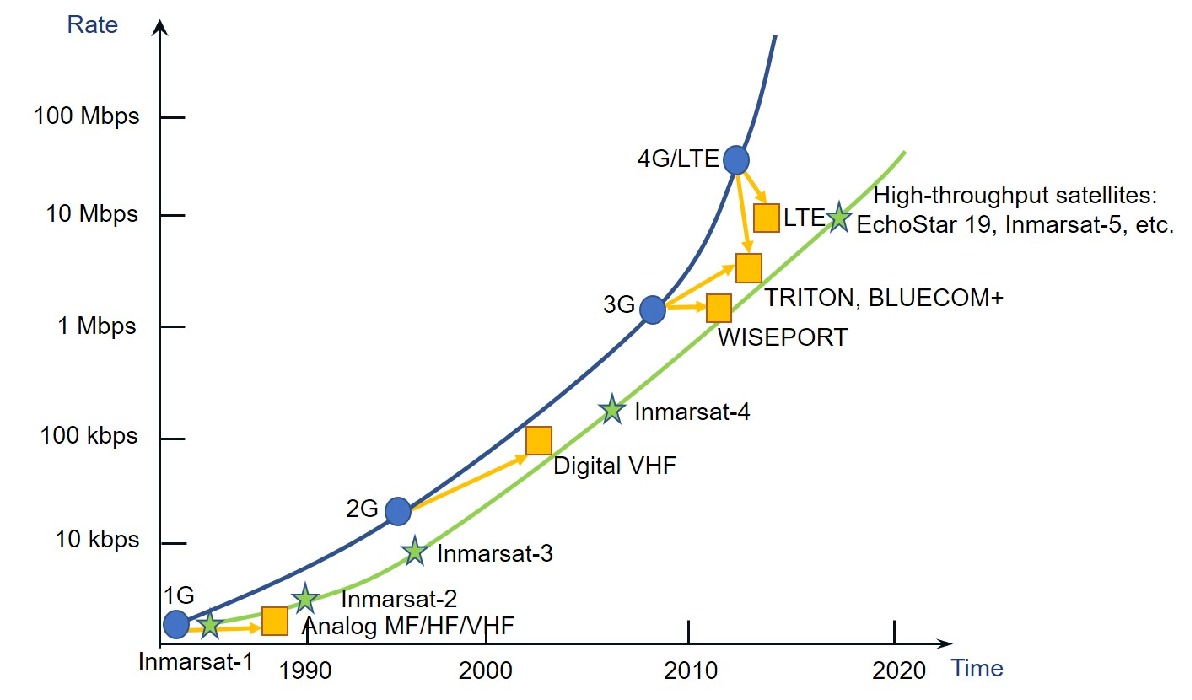} %这行为导入图片文件名称
\end{center}
\vspace*{-2mm}
\caption{The developing route of MCNs.}\label{fig:2}  %这行为文章中显示图片的标题
%\vspace*{-2mm}
\end{figure*}

The history of the development of MCNs is depicted in Figure 2.
Based on the network architecture, MCNs can be categorized into satellite-based, shore-based, island-based, vessel-based, and air-based networks. They will be discussed in the following sections.

\subsection{Satellite-based Maritime Communications}

%1) Inmarsat

Inmarsat is an international geostationary Earth orbit (GEO) satellite communications system.
It aims to provide worldwide voice and data services for various applications, such as ocean transport, air traffic control, and emergency rescue \cite{p1305}.
%Its main users are maritime applications and can provide voice and low-speed data services for them.
The first generation of Inmarsat systems (Inmarsat-1) was put into use in 1982. The system is composed of several satellites and transponders rented from other companies and organizations, mainly providing analogue voice, fax, and low-speed data services \cite{p1301}.
The second-generation system (Inmarsat-2) was put into use in 1990. It has a total of four satellites, each of which is equipped with a single global beam, providing digital voice, fax, and low/medium-speed data services \cite{p1303}\cite{p1309}.
The third-generation system (Inmarsat-3) was put into use in 1996. There are 5 satellites, and each satellite has 4--6 regional spot beams in addition to the global beam. Inmarsat-3 can support mobile packet data service (MPDS), with a capacity 8 times that of Inmarsat-2 \cite{p1306}--\cite{p1302}.
The state-of-the-art Inmarsat-4 system consists of 3 satellites. Each satellite has a global beam, 19 regional beams, and approximately 200 narrow spot beams. Inmarsat-4 can provide Internet services with a peak rate of 492 kbps \cite{p1307}\cite{p1310}.
The future Inmarsat-5 system, also known as Global Xpress, aims to provide worldwide customers with downlink services at 50 Mbps and uplink services at 5 Mbps \cite{p208}.

%2) Iridium

O3b is the first medium Earth orbit (MEO) satellite communications system that has been commercially used.
It consists of 16 active satellites, providing standard and limited services for areas within latitudes of 45 degrees and 62 degrees, respectively.
At present, the O3b company is actively promoting maritime satellite communication services and has installed O3b satellite communication terminals on several Royal Caribbean cruise ships. The maximum data rate of a single ship is 700 Mbps, while the delay is approximately 140 ms \cite{p10001}.

Iridium is an LEO satellite communications system providing voice and low-speed data services for users with satellite phones and pagers.
The second generation of Iridium satellite constellations, Iridium Next, started in 2017.
It consists of 66 active satellites, 9 in-orbit spare satellites, and 6 on-ground spare satellites.
%The Iridium-Next system aims to be a supplementary means to the terrestrial cellular communication system.
At present, Iridium Next provides data services of up to 128 kbps to mobile terminals and up to 1.5 Mbps to Iridium Pilot marine terminals.
In the future, it will support more bandwidth and higher rate, reaching a transmission rate of 1.4 Mbps for mobile terminals and up to 30 Mbps for high-speed data services when large user terminals are available \cite{p1401}.

%3) Tiantong No. 1

Tiantong-1 is China's first mobile satellite communications system, which is also known as the Chinese ``Inmarsat''.
The system was launched into orbit in 2016 and put into commercial use in 2018.
It mainly covers the Asia-Pacific region, including most of the Pacific Ocean and the Indian Ocean.
It provides voice, short message, and low-speed data services, with a peak rate of 9.6 kbps \cite{p2304}.

The Shijian-13 communications satellite is China's first high-throughput communication satellite. It is a multi-beam broadband communication system using the Ka-band, and its total communication capacity is more than 20 Gbps, approximately 10 times higher than before. The satellite is designed with 26 user spot beams, covering nearly 200 km of China's offshore areas \cite{p2305}.

Another high-throughput satellite EchoStar-19 has a capacity of more than 200 Gbps and is equipped with 138 customer communications beams and 22 gateway beams. The satellite provides users in North America with high-speed Internet services and emergency response. In addition, Ka-band-based airborne broadband services will be available on the EchoStar-19 satellite \cite{p2306}.

The satellite-based communication systems have wide coverage and can provide low-speed or high-speed data services depending the bandwidth.
However, satellite-based communications are easily affected by climate and the marine environment, resulting in low reliability \cite{p714}--\cite{p718}. In addition, the cost of ship-borne equipment and the communication charges are also very high. For example, the cost of installing ship-borne equipment for Inmarsat (Fleet 77) is approximately \$28000, including the antenna, terminal, handset, manuals, SIM card and power supply, and the data service costs \$2.8 per minute \cite{p1501}.
Data from the AIS show that there are nearly 80,000 ships sailing simultaneously around the world, less than 25,000 of which are high-end ships (with a load of more than 10,000 tons) that may afford the ship-borne equipment for high-throughput satellite communications.

%\subsection{Satellite-based Maritime Communications}

\subsection{Shore-based Maritime Communications}

%4) Japanese lighthouse visible light communication
%surface lighthouse long distance visible light communication.

%8) NAVTEX

The NAVTEX system is a narrow-band system with data rates of 300 bps, providing direct-printing services for ships within 200 nautical miles offshore.
It operates at the MF band, using the 518 kHz band to broadcast international information and the 490 kHz band for local messages \cite{p2301}.
The NAVTEX system delivers navigational messages, meteorological warnings and forecasts and emergency information to enhance marine safety,
%the NAVTEX system in other languages operates in the 490 kHz band.
but it cannot provide broadband communication services or obtain real-time information from users \cite{p1027}.

%9) PACTOR

The PACTOR system is also a narrow-band system, which operates at the HF band, using frequencies between 1 MHz and 30 MHz \cite{p2302}.
The first generation of PACTOR (PACTOR-I) was built to provide a combination of direct-printing and packet radio services.
Adaptive modulation methods and orthogonal frequency division multiplexing (OFDM) technologies were applied to
PACTOR-II and PACTOR-III, respectively, in order to improve the spectral efficiency \cite{p2303}.
%In 2010, the International Telecommunication Union Radiocommunication Group (ITU-R) proposed three HF radio systems and data transmission protocols for maritime communications, all of which use orthogonal frequency division multiplexing to improve .
%One of the most important systems is .
%The typical coverage of the system is 4000--40000 km, with data rates of 9.6 kbps and 14.4 kbps.
The state-of-the-art PACTOR-IV system uses adaptive channel equalization, channel coding, and source compression techniques, and has proven to be suitable for channels with severe multi-path.
%to achieve twice the data rate using the same power and bandwidth as PACTOR-3.
%Another important system based on this protocol is IPBC (Internet protocol for boat communications). The IPBC system covers the offshore range (40 to 250 nautical miles) using the lower frequency band (48 MHz), and the offshore area over 200 nautical miles using higher frequency band (826 MHz).
%Unfortunately,
PACTOR-IV can provide text-only e-mail services for ships thousands of kilometres away from the land with a data rate of up to 10.5 kbps, using a bandwidth of 2.4 kHz \cite{p10002}.
Similar to NAVTEX, the PACTOR system cannot provide real-time communication services due to a large transmission delay.

%In October 2008, the Visible Light Communication Coalition (VLCC) of Japan implemented visible light communications at the maritime lighthouse. The communication distance was 2 km and the speed was 1022 bps. In 2009, Japan's Outstanding Technology Corporation developed a visible light communication system that has successfully achieved 13 kbps of analog voice signal transmissions. Visible light communications are less vulnerable to interference. In the harsh environment at sea, the visible light communications can provide secure communication services within the line-of-sight (LOS).
%LTE-based projects

As wireless communications technologies advance, several broadband MCNs have been constructed.
The world's first offshore LTE network
%The project
was jointly developed by Tampnet in Norway and Huawei in China.
%It is the world's first network application that integrates LTE technology with off-shore communications,
It covers the platforms, tankers, and floating production storage and offloading (FPSO) facilities from 20 km to 50 km offshore on the North Sea, providing voice and data services of 1 Mbps uplink and 2 Mbps downlink.
It also supports video surveillance data uploading and wireless trunking services \cite{p2308}.

Ericsson has also been working on connecting vessels at sea with shore-based BSs. It aims to enable maritime services that facilitate crew infotainment, cargo monitoring, and shipping route optimization.
Ericsson and China Mobile have constructed a TD-LTE trial network for maritime coverage in Qingdao, China.
The network operates at the 2.6 GHz band, covering areas up to 30 km offshore with a peak rate of 7 Mbps.
It can provide broadband services for offshore applications, such as maritime transportation and offshore fisheries \cite{p2309}.

The shore-based MCNs, as extensions of terrestrial networks, can provide broadband communications services for offshore applications, such as multimedia file downloading and video surveillance data uploading.
However, the shore-based MCNs have limited coverage compared with satellite networks, and the coverage performance depends largely on the geometrically available BS sites. Shore-based communications are suitable for maritime applications that are densely clustered in a small area.

\subsection{Island-based Maritime Communications}

For the remote islands on the sea, high-quality communications can not only provide service for the islanders but also provide strong support for the timely communications and interconnection of border information.
In 2015, the U.S. wireless provider Verizon Communications enhanced 4G LTE network coverage on Rhode Island. It can provide the islanders and nearby vessels with web browsing and file downloading services \cite{p2311}.
In 2016, China Mobile set up a 4G BS on the Yongshu Reef, which is more than 1,400 km from mainland China.
By building satellite ground stations on the island, the signals from the island can be transmitted to the satellites, then to the backbone gateway on the mainland. The transmission rate often reaches 10 Mbps on the island and 15 Mbps using nearby ship-borne communication equipment.
In 2017, China Telecom set up four 4G BSs on the Nansha Islands, which were connected to the mainland using underwater cables.
The BSs provide coverage for the islands and reefs such as Yongshu Reef, Qibi Reef, Meiji Reef and nearby sea areas, enhancing broadband communication services \cite{p2310}.
%In 2018, the 4G BSs on Kaishan Island in China came into service.
%Field tests have shown that the downlink rate averages over 80 Mbps, which can

The construction of island-based BSs further expands the coverage of terrestrial mobile signals. Island-based MCNs can support clear voice and video calls from the coast to the island and provide high-quality communication services for the surrounding ships and fishermen. On the other hand, island-based BSs are more vulnerable to extreme climate events, such as typhoons and rainstorms. Their coverage is also limited.

\begin{table*}
\centering
\caption{Existing Maritime Communication Networks.}
\scriptsize
\begin{tabular}{|c|c|c|c|c|c|}  %通过添加|来表示是否需要绘制竖线
\hline  % 在表格最上方绘制横线
\textbf{Project/System} & \textbf{Sponsor} & \textbf{Frequency} & \textbf{Maximum rate} & \textbf{Coverage} & \textbf{Feature}\\
\hline  %在第一行和第二行之间绘制横线
WISEPORT & Singapore & 5.8 GHz & 5 Mbps & 15 km & WiMAX\\
\hline  %在第一行和第二行之间绘制横线
TRITON & Singapore & 5.8 GHz & 6 Mbps & 27 km & mesh\\
\hline  %在第一行和第二行之间绘制横线
Maritime-MANET & Japan & 27/40 MHz & 1.2 kbps & 70 km & Ad Hoc\\
\hline  %在第一行和第二行之间绘制横线
BLUECOM+ & Portugal	& 500/800 MHz & 3 Mbps & 100 km & balloons, 2-hop \\
\hline  %在第一行和第二行之间绘制横线
Digital VHF TMR	& Norway & 87.5-108/174-240 MHz	& 21/133 kbps & 130 km & broadcasting\\
\hline  %在第一行和第二行之间绘制横线
Qingdao TD-LTE Trial Network	& China Mobile \& Ericsson & 2.6 GHz	& 7 Mbps & 30 km &  LTE \\
\hline  %在第一行和第二行之间绘制横线
Norwagian Offshore LTE Network & Tampnet \& Huawei & 1785-1805 MHz& 2 Mbps	& 50 km & LTE \\
\hline  %在第一行和第二行之间绘制横线
Internet.org	& Facebook & laser & unknown & wide &  UAVs \& laser\\
\hline  %在第一行和第二行之间绘制横线
Loon	& Google & 2.4 GHz	& 10 Mbps & wide &  balloons \\
\hline  %在第一行和第二行之间绘制横线
Inmarsat-4	& Inmarsat & L/S band & 492 kbps & wide & GEO \\
\hline  %在第一行和第二行之间绘制横线
Iridium NEXT & Iridium \& Motorola &  L band & 30 Mbps & wide & LEO \\
\hline  %在第一行和第二行之间绘制横线
Tiantong-1 & China & S band	& 9.6 kbps & wide & GEO \\
\hline % 在表格最下方绘制横线
\end{tabular}
\end{table*}

\subsection{Vessel-based Maritime Communications}

The Japanese Ministry of Internal Affairs and Communications has developed a maritime mobile ad hoc network (Maritime-MANET) to expand the coverage of shore/island-based MCNs via ship-to-ship communications.
The network uses 27 MHz and 40 MHz frequency bands, covering areas of up to 70 km offshore.
However, the supported rate is only 1.2 kbps, supporting mainly narrowband communication services, such as the short message service (SMS) \cite{p2307}.

Singapore has launched the TRITON project to develop a wireless mesh network to expand the coverage area.
In this network, each vessel, maritime beacon, or buoy serves as a mesh node, which can route traffic for other nearby nodes.
The network operates at the 5.8 GHz band, covering areas up to 27 km away from the shore, with a coverage of 98.91\%.
%In this network, the near-shore vessels form a multi-hop mesh network and access the terrestrial broadband communication network through the on-shore BS.
It can provide broadband communication services of 6 Mbps for offshore applications but cannot cover the high seas \cite{p204}.

The vessel-based mesh or ad hoc networks can provide broadband communication services for most vessels and platforms along the coast.
However, the link stability of vessel-based MCNs is restricted by the frequent change of the sea surface and marine weather conditions.
In addition, the mesh architecture requires a high density of vessels, and each vessel needs to install expensive equipment.
Therefore, more reliable network protocols and more cost-effective ship-borne terminals are required for vessel-based maritime communications.

\subsection{Air-based Maritime Communications}

%5) Internet.org project

%Facebook is near to the space long drone communication system.

The Internet.org project was launched by Facebook in 2013, aiming to provide free Internet access for users in remote areas, including marine users.
The project utilizes UAVs at altitudes of 55--82 km to serve as aerial BSs and form a network via laser communications.
Until now, Facebook has teamed up with a set of mobile operators and handset manufactures and has found a number of sites for deploying UAVs to cover impoverished areas in Latin America, Asia and Africa \cite{p205}.

%6) Project Loon

The Loon project was initiated by Google in 2013, aiming to provide Internet access for users in the countryside and remote areas.
The project uses super-pressure balloons at an altitude of 20 km to build a communication network. The network operates at the 2.4/5.8 GHz band and can provide communication services of 10 Mbps. Although the project is not commercial yet, it has provided emergency communication services for several areas suffering from natural disasters \cite{p206}.

The BLUECOM+ project also uses tethered balloons as routers to extend land-based communications to remote ocean areas. It exploits the TV white spaces for long-range wireless communications and uses multi-hop relaying techniques to extend the coverage. Simulation results have shown that the BLUECOM+ solution can cover the ocean areas up to 150 km from shore, providing broadband communication services at 3 Mbps \cite{p201}\cite{p2003}.

In general, the air-based MCNs can cover a wider area than the vessel-based MCNs, as the BSs are high above the ocean surface.
They can provide remote users with high-rate and low-reliability communication services.
On the other hand, aerial BSs, such as UAVs and balloons, are easily damaged by severe weather.
%\textcolor{black}{\subsection{Underwater Communications}}
%
%\textcolor{black}{
%The most representative of the underwater acoustic communication network is the US Seaweb network. The network is designed to verify the performance of underwater acoustic communication networks in shallow sea complex channels. Since the first ocean trials in Buzed Bay, Massachusetts, in 1998, more than 50 sea trials have been conducted. Seaweb demonstrates excellent performance in applications such as underwater mobile platform command transmission, underwater positioning navigation, and ocean observation \cite{p10401}.
%}
%
%\textcolor{black}{
%Russia completed the construction of the underwater GLONASS positioning system in 2018. The system consists of a sonar buoy equipped with satellite communications and ultrashort wave radios and an underwater unmanned aerial vehicle. The system is positioned to a depth of up to 8 km underwater and can serve underwater oil and gas exploration \cite{p10402}.
%}
%
\textcolor{black}{\subsection{Sensing-Oriented Maritime IoT}}
\textcolor{black}{DARPA launched the Maritime IoT project in 2017, which plans to deploy tens of thousands of small, low-cost smart floating objects to form a distributed sensor network to achieve continuous situational awareness of large areas of the sea. Each smart float will use a set of commercial sensors to collect environmental data such as sea temperature, sea conditions and location in the area, as well as activity data for commercial vessels, aircraft and even marine animals. These floats can periodically transmit data via satellite to the cloud for storage and real-time analysis. The first phase of the Maritime IoT mainly involves the initial designs and trials to verify concepts \cite{p10403}.}

%\subsection{Ship-borne BS based Maritime Communications}

The key performance indicators of existing MCNs are compared in Table II.
%As can be seen from the above,
For satellite communications, narrow-band satellites mainly provide services, such as telephone, telegraph and fax, and the communication rate is low.
The newly launched high-throughput satellites enable broadband maritime coverage.
However, the cost of ship-borne equipment is very high.
%High-throughput satellites are not cost-effective and not suitable for providing offshore broadband information services.
In addition to satellites, shore \& island-based BSs can be built to extend the maritime coverage of terrestrial networks. UAVs, high-end ships and offshore lighthouses can be exploited as well to extend the coverage further.
The coverage performance of the MCN depends largely on the abovementioned geometrically available BS sites, and
%In terms of shore-based communications, NAVTEX, AIS and other systems mainly provide services such as information broadcasting, voice and ship identification using IF/HF/VHF bands, but cannot provide high-speed data services.
%LTE-based maritime trial networks mainly generate directional beams through array antennas for fixed-point coverage, but do not support large-scale mobile communications.
the transmission efficiency of a single BS is affected by maritime weather conditions, e.g., wave fluctuations. The link stability is generally poorer than terrestrial networks.
%Therefore,
\textcolor{black}{From the above, it is necessary to enhance the transmission efficiency in the complex and dynamical maritime environment, to extend the coverage by taking advantage of different methods, and to develop service-oriented transmission and coverage techniques to meet the unique service requirements from maritime applications. We begin with the key technologies for transmission efficiency enhancement in the following section.}
\\

%\begin{figure} [htb]
%\begin{center}
%\includegraphics*[width=10cm]{fig20181006.eps} %这行为导入图片文件名称
%\end{center}
%\vspace*{-4mm} \caption{Existing MCNs: Broadband MCNs with wide coverage are needed.}\label{fig:2}  %这行为文章中显示图片的标题
%\vspace*{-2mm}
%\end{figure}

\section{Enhancing Maritime Transmission Efficiency}\label{sec:3}
%\subsection{Challenge: Unique channel characteristics}

Compared with the terrestrial environment, the atmosphere over the sea surface is unevenly distributed due to seawater evaporation. Thus, the electromagnetic propagation over sea is susceptible to sea surface conditions (tidal waves, etc.) and atmospheric conditions (temperature, humidity, wind speed, etc.), as depicted in Figure 3.
In addition, the height and angle of ship-borne antennas can change rapidly with the fluctuation of the sea surface.
A representative insight on this issue can be found in \cite{p10401}, where the impact of sea waves to radio propagation and the communications link quality has been comprehensively discussed.
Hence, maritime channel fading is particularly sensitive to parameters such as antenna height and angle,
which may cause frequent link interruption.
% and lower the transmission efficiency.
%Therefore,
These complex time-varying factors reduce the transmission efficiency in maritime scenarios.
% is always low affected by .

To enhance the transmission efficiency, it is necessary to understand and take advantage of the characteristics of the wireless propagation environments over sea. Compared with terrestrial scenarios, electromagnetic propagation over sea is affected by various weather conditions, such as sea surface conditions and atmospheric conditions. These are unique for maritime channels. Therefore, the measurement and modelling of the maritime channel is very important for the design of MCNs \cite{p304}.
On the other hand, advanced resource allocation schemes, such as dynamic beamforming and user scheduling techniques, need to be studied to utilize the dynamic changes of maritime channels.
In addition, evaporation ducts may exist due to uneven atmospheric humidity above the sea surface, which can trap the signal inside and greatly reduce the transmission loss. It is possible to exploit the evaporation duct effect to improve the transmission efficiency, especially for remote transmissions.
We start with the characteristics and models for maritime channels.

\begin{figure} [t]
\vspace*{1mm}
\begin{center}
\includegraphics*[width=9cm]{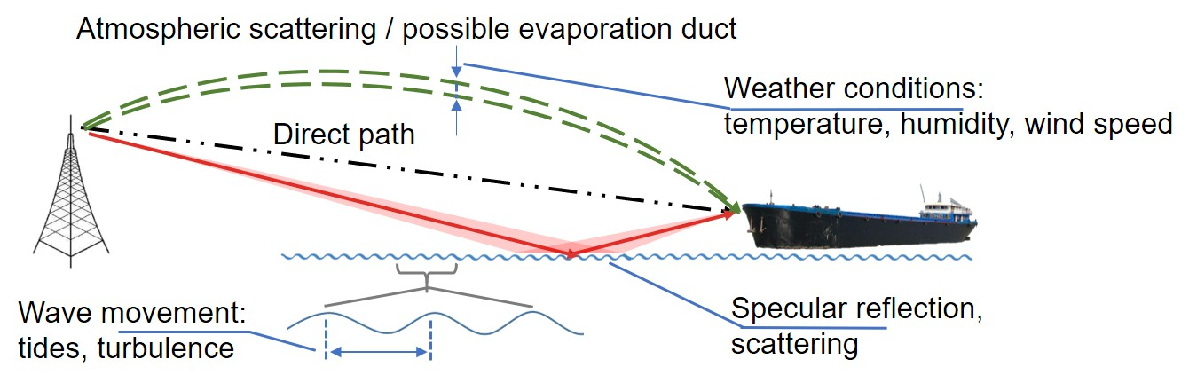} %这行为导入图片文件名称
\end{center}
\vspace*{-1mm}
\caption{Illustration of typical shore-to-ship propagation rays, which are affected by sea state and atmosphere conditions.}\label{fig:2}  %这行为文章中显示图片的标题
%\vspace*{-2mm}
\end{figure}

\subsection{Characteristics and Models of Maritime Channels}

%\subsection{Maritime Channel Measurement and Modelling Projects}

%At present,
%A number of universities and research institutes in the United States, South Korea, Singapore, Norway, and other countries have carried out maritime channel measurement and modelling work.
Various channel measurements and modelling works have been conducted to analyse the impact of system parameters (frequency, antenna height, etc.) and maritime environments (sea state, weather conditions, etc.) on maritime channel fading \cite{p306}-\cite{p1202}.
%, and have achieved some phased results.
%According to the structure of maritime channel model, we classify the large-scale CSIT and the small-scale CSIT
%The large-scale parameters vary slowly.
According to the rate of change of these parameters over time, maritime channel fading can be classified as large-scale fading and small-scale fading. The large-scale fading varies slowly on the same order as the user location change. The small-scale fading is much faster due to the rapid fluctuations in signal amplitude, phase, or multi-path delay over a signal wavelength.
%The multi-path interference effect caused by sea surface reflection, refraction and scattering is an important factor that affects the small-scale fading.

For the large-scale fading, Y. Bai \emph{et al.} studied the influence of ground curvature on the signal propagation characteristics of the maritime environment and calculated the link budget for Wideband Code Division Multiple Access (WCDMA) systems \cite{p306}. K. Yang \emph{et al.} studied the possibility of adapting several terrestrial channel models to the maritime environment in the 2 GHz band and found that the model from the International Telecommunication Union Radiocommunication Group (ITU-R) agrees well with the measurement results \cite{p312}. However, the ITU-R model uses simple corrections for different terrains and therefore does not truly reflect the complex maritime variations such as sea reflections and evaporation ducts.

Considering the impact of sea surface reflections, some recent works \cite{p302}--\cite{p311} studied the two-ray channel model and proposed several modified models. Among them, Y. Zhao \emph{et al.} considered factors, such as sea surface reflection and antenna height, and proposed a two-ray model suitable for maritime channels \cite{p302}.
The model assumes that the maritime channel mainly consists of a direct path and a reflection path, and its path loss can be expressed as
\begin{align}
\!\!{L_{2{\rm{ - ray}}}} =  - 10{\log _{10}}\left\{ {{{\left( {\frac{\lambda }{{4\pi d}}} \right)}^2}{{\left[ {2\sin \left( {\frac{{2\pi {H_1}{H_2}}}{{\lambda d}}} \right)} \right]}^2}} \right\}\!\!
\end{align}
where ${\lambda }$ is the carrier wavelength, ${d}$ is the distance between the transmitting antenna and the receiving antenna, ${H}_{1}$ and ${H}_{2}$ represent the heights of the transmitter antenna and the receiver antenna, respectively.

Additionally, J. C. Reyes-Guerrero \emph{et al.} measured the maritime channel in non-line-of-sight (NLOS) scenarios and proposed a simplified two-path model by using a geometrical approximation method. Compared with the free-space model and the two-ray model, this model is only appropriate for transmission over a short distance \cite{p301}. N. Mehrnia \emph{et al.} introduced the index correction coefficient in the two-ray model formula and obtained better channel prediction in the 5 GHz band \cite{p315}. Jae-Hyun Lee \emph{et al.} studied large-scale fading characteristics and small-scale fading characteristics in the 2.4 GHz band and found that the two-ray model considering the wave height is more consistent with the experimental data in general \cite{p311}. This modified two-ray model can achieve good accuracy under certain scenarios but is only applicable to offshore areas within a short distance.

%K. Yang et al. measured the channel between the transmitting antenna on the far sea and the receiving antenna on the shore, and analysed the important influence of the antenna position on signal propagation based on the received signal level and the power delay spectrum [9]; Y. Bai et al. studied the influence of ground curvature on the signal propagation characteristics of sea areas, and conducted link budgeting for WCDMA systems [10]. Y. Zhao et al. considered factors such as sea surface reflection and antenna height, and proposed a two-path model suitable for the maritime channel.

%In addition, J. C. Reyes-Guerrero et al. measured the maritime channel under non-LOS scenarios and compared it with the free-space model and the two-path model [12]. Y.H Lee et al. measured the near-shore channel under the LOS scenario and considered the evaporation duct effect caused by the uneven water vapor density above the sea surface. Based on the two-path model, a three-path model was proposed. As shown, the refraction diameter is affected by the antenna height of the transmitting and receiving antenna, as well as various weather factors such as sea temperature and tidal effects [13][14].

In the marine atmosphere, special atmospheric refractive index structures easily form evaporation ducts, so that the electromagnetic wave has an extra scattering energy gain, enabling it to propagate to more distant areas. The evaporation duct effect is necessary for communications at a longer distance. Y. H. Lee \emph{et al.} measured the near-shore channel in the line-of-sight (LOS) scenario. The analysis shows that, when the distance between the transmitter and receiver exceeds a certain threshold (relative to the antenna heights), the presence of the evaporation duct will affect the path loss. In addition, Y. H. Lee \emph{et al.} proposed a three-ray path loss model, which is closely related to the heights of the evaporation duct and the transmitting and receiving  antennas \cite{p303}. The height of the evaporation duct can be estimated using the Paulus-Jeske empirical model (P-J model). A. Coker \emph{et al.} simulated and analysed the effect of evaporation duct height on signal attenuation and diversity \cite{p317}. More recently, in \cite{p6015}, the authors proposed a way to estimate the evaporation duct height using a novel refractivity profile model. Under proper sea conditions, the 3-ray model considering the evaporation duct has considerable advantages over the 2-ray model, and its path loss can be expressed as
\begin{align}
\!\!\!\!{L_{{\rm{3 - ray}}}} \!\!=\!\! \left\{\!\!\!\! \begin{array}{l}
 - 10{\log _{10}}\! \left\{ {{{\left( {\frac{\lambda }{{4\pi d}}} \right)}^2}{{\left[ {2\sin \left( {\frac{{2\pi {H_1}{H_2}}}{{\lambda d}}} \right)} \right]}^2}} \right\}, d \!<\! {d_{break}}\\
 - 10{\log _{10}}\! \left\{ {{{\left( {\frac{\lambda }{{4\pi d}}} \right)}^2}{{\left[ {2\left( {1 + \Delta } \right)} \right]}^2}} \right\}, d \!>\! {d_{break}}
\end{array} \right.\!\!\!\!\!\!\!\!\!\!
\end{align}
where $\Delta  = 2\sin \left( {\frac{{2\pi {H_1}{H_2}}}{{\lambda d}}} \right)\sin \left( {\frac{{2\pi \left( {{H_e} - {H_1}} \right)\left( {{H_e} - {H_2}} \right)}}{{\lambda d}}} \right)$, ${d_{break}}{\rm{ = }}\frac{{4{H_1}{H_2}}}{\lambda }$, and ${{H_e}}$ denotes the height of evaporation duct layer.

In addition to path loss, the maritime channel model needs to consider small-scale fading caused by sea-level fluctuation and atmospheric scattering. X. Hu \emph{et al.} pointed out that multi-path reflection on the sea surface can be divided into coherent specular reflection and non-coherent diffuse reflection, and the concept of effective reflection area was proposed \cite{p318}. M. Dong \emph{et al.} used the Rayleigh roughness decision criterion to prove that the diffuse reflection from the sea surface is negligible when the wave height is less than 4 metres and the grazing angle is less than 5 degrees \cite{p319}. K. Haspert \emph{et al.} proposed a theoretical approximation modelling method that can be applied to multi-path channels containing specular and diffuse reflection components \cite{p320}. K. Yang \emph{et al.} measured the channel between the transmitting antenna on the far sea and the receiving antenna on the shore, and analysed the important influence of the antenna position on signal propagation based on the received signal level (RSL) and the power delay profile (PDP) \cite{p305}. J. Lee \emph{et al.} analysed the probability density function (PDF) of the small-scale fading and pointed out that the PDF is more approximate to the Rice distribution than the Nakagami-m distribution and the Rayleigh distribution \cite{p307}. K. Yang \emph{et al.} analysed the Doppler shift \cite{p308}. F. Huang \emph{et al.} considered the smooth sea surface and the rough sea surface.
The impulse response of a multi-path channel, composed of the direct path, reflected paths, and scattering paths, was obtained. The model is suitable for different carrier frequencies, transmission distances, and sea states \cite{p324}.
More recently, in \cite{p6021}, the authors performed ship-to-shore propagation measurements at the 1.39 GHz band and the 4.5 GHz band, and proposed a model to capture the behaviour of small-scale fading at different frequency bands.

Focusing on the influence of various factors such as waves, tides, and evaporating ducts on the maritime channel, we conducted a maritime channel measurement experiment at the 5.8 GHz band on the East Sea of China. The bandwidth of the measured signal is 20 MHz and the maximum distance is 33 km.
The transmitter is set at the top of the teachers' apartment of the Qidong Campus of Nantong University, and the height is approximately 22 m. The receiver is arranged on the top of the fishing boat cabin and the height is approximately 4 m. The vessel travels in a straight line in the East China Sea to the east at a constant speed of 10 knots.
Parameters that affect the large-scale channel fading include the carrier frequency, the antenna heights, and the distance between the transmitting antenna and the receiving antenna.
In particular, in the maritime environment, the height of the wave changes slowly due to the tide phenomenon, which changes the height of the ship-borne antennas consequently \cite{p913}. It affects the received signal strength in duplicate measurements according to the two-ray model, as shown in Figure 4.
The small-scale fading characteristics of the channel can be observed by deriving the probability density distribution from the measurement data of the path loss, which can be
%The probability density distributions commonly used to describe small-scale fading are Rayleigh distribution and Rician distribution. %Figure 6 shows the path loss deviation distribution obtained by channel measurement, and uses the Rician distribution and the Rayleigh distribution to fit the results, respectively.
described by the Rician distribution in LOS due to the existence of direct path and multiple sea surface reflection paths. However, the measurement results deviate greatly from the Rician distribution, and the small-scale fading model of the maritime channel needs further exploration.

\begin{figure} [t]
\begin{center}
\includegraphics*[width=9cm]{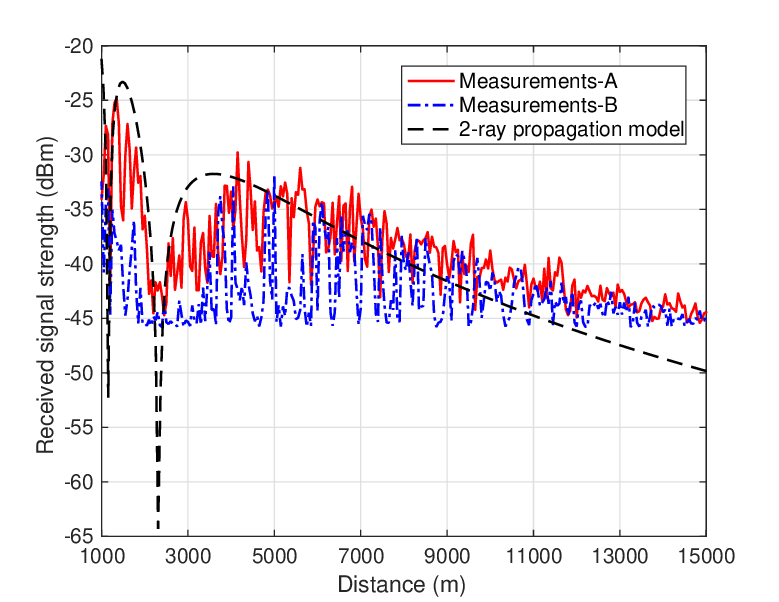} %这行为导入图片文件名称
\end{center}
\vspace*{-1mm}
\caption{Received signal strength results of channel measurements conducted on East China Sea and a two-ray channel model.}\label{fig:2}  %这行为文章中显示图片的标题
%\vspace*{-2mm}
\end{figure}

For satellite channels, the channel fading consists of free space loss, ionospheric scintillation effect, atmospheric absorption loss, multi-path fading and shadowing effects. When the weather is good, the signal is not blocked by clouds when it is transmitted over the channel. The signal received by the terminal includes scattering and direct components. In this case, the received signal envelope follows the Rice distribution.
%, i.e.,
%\begin{align}
%p\left( a \right) = \frac{a}{{{\sigma ^2}}}\exp \left( { - \frac{{{a^2} + {\rho ^2}}}{{2{\sigma ^2}}}} \right){I_0}\left( {\frac{{a\rho }}{{{\sigma ^2}}}} \right).
%\end{align}
When the weather conditions are poor, the signal is affected by both shadowing effect and multipath without direct signal. It can be described using the Suzuki model as
\begin{align}
\!\!\!\!p\left( a \right) \!=\!\!\! \int_0^\infty \!\!\!\! {\frac{a}{{{\sigma ^2}}}\exp\!  \left(\! { - \frac{{{a^2}}}{{2{\sigma ^2}}}}\! \right)\!\frac{1}{{\sqrt {2\pi } \sigma {\sigma _l}}}} \!\exp\!  \left(\! { - \frac{{{{\left( {\ln \sigma \! -\! \mu } \right)}^2}}}{{2\sigma _l^2}}} \!\right)\!d\sigma
\end{align}
where $\sigma$ is the standard deviation of each Gaussian component, $\mu$ and ${\sigma _l}$ are the mean and standard deviations of signals that follow the Log-normal distribution, respectively.
For maritime satellite channels, the authors in \cite{p330} measured the channel fading with different antenna types and
elevation angles and compared the performance of several modulation schemes. The authors in \cite{p331} and \cite{p332} analysed the characteristics of rain fading on the Ka-band using the statistics extracted from the satellite-to-beacon propagation measurements.

\newcommand{\tabincell}[2]{\begin{tabular}{@{}#1@{}}#2\end{tabular}}%放在导言区
\begin{table*}
\centering
\caption{Maritime Channel Measurements and Models.}
\scriptsize
\begin{tabular}{|c|c|c|c|c|c|c|c|}  %通过添加|来表示是否需要绘制竖线
\hline  % 在表格最上方绘制横线
\textbf{Ref.} & \textbf{Scenario} & \tabincell{c}{\textbf{Frequency}\\\textbf{(GHz)}} & \tabincell{c}{\textbf{Tx-Rx }\\\textbf{Distance (km)}} & \tabincell{c}{\textbf{Tx/Rx Antenna }\\\textbf{Heights (m)}} & \textbf{Channel Statistics} & \tabincell{c}{\textbf{Environmental}\\ \textbf{Factors Considered}} & \textbf{Channel Model} \\
\hline  %在第一行和第二行之间绘制横线
\cite{p306} & ship to shore	& 2.1 & 13-41	& 10,25,50,100/10 & PL & earth curvature & FSPL model\\
\hline  %在第一行和第二行之间绘制横线
\cite{p312} & ship to shore & 2.075 & 45 & 9.5/11.2 & RSL, PDP, SC  & not mentioned & ITU-R model \\
\hline % 在表格最下方绘制横线
%\cite{p302} & simulation &1.8 & 50-90 & 200/10 & Tx loss & weather, sea state & ITM \\
%\hline % 在表格最下方绘制横线
\cite{p301} & buoy to ship & 5.8 & 10 & 1.7/9.8 & PL & not mentioned & modified 2-ray model \\
\hline % 在表格最下方绘制横线
\cite{p315} & ship to ship & 35/94 & 20 & 5/9.7 & PDP, PL, RMS-DS & not mentioned &  FSPL model, modified 2-ray model\\
\hline % 在表格最下方绘制横线
\cite{p311} & shore to ship & 2.4 & 2 & 3/4.5 & RSL, PL & not mentioned & FSPL, ITU-R, and 2-ray models\\
\hline % 在表格最下方绘制横线
\cite{p303} & ship to shore	& 5.15	&10	&3-4/7.6,10,20 & PL	& evaporation duct & FSPL, 2-ray, and 3-ray models \\
\hline % 在表格最下方绘制横线
\cite{p305} & ship to shore	& 2.075	& 45 & 9.5/11.2 & RSL, PDP, SC & not mentioned & ITU-R model\\
\hline % 在表格最下方绘制横线
\cite{p307} & air to ground & 5.7 & 10 & 370,1830/2.1,7.65 & PL & evaporation duct & FSPL model, 2-ray model\\
\hline  %在第一行和第二行之间绘制横线
\cite{p308} & ship to shore & 2.075	& 15.5 & 6.5/23	& RSL, PDP, DFO, SRC & sea state & 2-ray model \\
\hline % 在表格最下方绘制横线
\cite{p309} &shore to ship & 1.95 & 16 & 22/2.5 & RSL, PL & not mentioned & FSPL model\\
\hline % 在表格最下方绘制横线
\cite{p310} & flight to ship & 5.7	& 27.7 & 1000/5.5 & PL	& evaporation duct & ducting-induced enhancement model\\
\hline % 在表格最下方绘制横线
\cite{p329} & island to island & 0.248/0.341 & 33.3,48 & 18.5/16,14 & RSL, PL & sea state & FSPL model, ITU-R model\\
\hline % 在表格最下方绘制横线
\cite{p327} & buoy to ship & 5800 & 0.2 & 1.9/3.3 & PDP, RMS-DS & not mentioned & not mentioned\\
\hline % 在表格最下方绘制横线
\cite{p328} & ship to ship & 1900 & 5-30 & 8/8 & PDP, PL & not mentioned & log-distance PL model\\
\hline % 在表格最下方绘制横线
\multicolumn{8}{|l|}{\tabincell{l}{Tx: transmitter; Rx: receiver; PL: path loss; RSL: received
signal level; PDP: power delay profile; SC: spatial correlation; RMS-DS: root-mean-square delay spread; \\ DFO: Doppler frequency offset; SRC: sea reflection coefficient; FSPL: free-space path loss}}\\
\hline % 在表格最下方绘制横线
\end{tabular}
\end{table*}

The channel parameters from representative maritime channel measurements and modelling works are listed in Table III.
The maritime channel model is determined not only by parameters such as signal frequency, transmission distance, antenna height and moving speed, but also by oceanic weather and sea surface fluctuations \cite{p333}--\cite{p1202}.
The above studies have considered several specific factors and measured the relevant received signal strengths under specific experimental setups and marine environments. However, their combined effect is still unknown. For the design of a practical MCN, link budget is necessary based on channel measurement results, which changes greatly from spring to winter, from day to night, and from sunny days to windy days. Therefore, network design ignoring the environmental factors will largely reduce the transmission efficiency and degrade the coverage performance of MCNs.
On the other hand, the transmission efficiency in MCNs is envisioned to be enhanced by using some promising 5G technologies, such as massive multiple-input multiple-output (MIMO) technologies, millimetre wave (mmWave) communications, and vehicle-to-vehicle (V2V) communications \cite{p3005}. Until now, many massive MIMO channel models \cite{p3008}--\cite{p3011}, mmWave channel models \cite{p3012}\cite{p3013}, V2V channel models \cite{p3014}\cite{p3015}, and high-mobility channel models \cite{p3016}--\cite{p3018} have been proposed, and a general 5G channel model can be used to simulate the channels \cite{p3006}.
However, these models are mostly based on the channel measurements in terrestrial scenarios and may not be suitable for the environment-sensitive maritime channels \cite{p3007}. Therefore, it is necessary to carry out further measurements and modelling studies on maritime channels.

%\begin{figure} [htb]
%\begin{center}
%\includegraphics*[width=9cm]{measure4.eps} %这行为导入图片文件名称
%\end{center}
%\vspace*{-4mm} \caption{Channel measurements conducted on East China Sea: Small-scale channel fading model is still undiscovered.}\label{fig:2}  %这行为文章中显示图片的标题
%\vspace*{-2mm}
%\end{figure}

\subsection{Reducing Transmission Loss: Exploiting Evaporation Duct for Remote Transmissions}

The atmospheric refractive index over the sea surface varies with the maritime environment.
Electromagnetic waves have different propagation paths depending on the rate at which the refractive index changes with height.
When the rate meets certain conditions, atmospheric ducts will be formed, and signals will be trapped therein, as depicted in Figure 5 \cite{p341}.
%For example, in the region where ?M/?h ≤ 0, the electromagnetic wave advances, which is called the atmospheric duct layer.
%Depending on the formation mechanism and the relationship between refractive index and height, atmospheric ducts can be classified into four categories: evaporation ducts, surface ducts containing base layers, surface ducts, and suspended ducts. Among them,
Atmospheric ducts can be utilized to improve transmission efficiency, as the propagation loss in the duct layer is much smaller than that in free space \cite{p343}.

Three types of atmospheric ducts often appear over the sea surface, namely, surface duct, elevated duct, and evaporation duct.
The evaporation duct, formed by a large amount of seawater evaporated approximately 0--40 m above sea level, is the most common type of atmospheric duct and only occurs in the oceanic atmosphere \cite{p341}. Using the evaporation duct, several radio links have been set up for beyond-LOS maritime communications, such as the 78-km link from the Australian mainland to the Great Barrier Reef \cite{p343}, and the 100-km link between Malaysian shores \cite{p344}\cite{p345}.
%The signal is bound in the evaporation duct layer, so the transmission loss in the evaporation duct layer is smaller and the transmission distance is longer than that of the free-space signal, enabling beyond-LOS propagation.
%However, current research work focuses on the application of atmospheric ducts in the radar field, while atmospheric ducts are less used in marine communications

It should be noted that the height of the evaporation duct layer depends on various environmental factors, such as air-sea temperature difference, humidity, air pressure, wind speed, and wave height \cite{p325}.
Although the P-J formulation can be used to calculate the duct height, it may lose the prediction accuracy due to its sensitivity to the weather information.
The utilization of evaporation duct for maritime communications is still in the early stage. To promote the development of MCNs, more meteorological instruments are needed to collect the vertical weather information, and more accurate models are required to predict the channel state information (CSI).

%Although
%The authors in \cite{p345} set up a microwave link between Australia and a reef 78 km away. The link operated at 10.6 GHz and provided a data rate of 10 MB/s.
%For the first time, high-capacity, long-distance over-the-horizon ocean communication was realized by using atmospheric ducts. %This paper further analysed the optimal frequency and optimal antenna height of atmospheric duct communication and concluded with the experimental tool AREPS (advanced refractive effects prediction system).
%Operating frequency 10.5 GHz, antenna height 4 m is most suitable for the realization of over-the-horizon maritime communications using evaporation ducts.
%In order to solve many problems of existing marine communication networks,
%The atmospheric duct effect is a new area in the research of marine communications. At present, the research on maritime communications based on atmospheric duct effect is still in its infancy, but the utilization of atmospheric duct effect is a promising way to promote the further development of maritime communications.

\begin{figure} [t]
\vspace*{1mm}
\begin{center}
\includegraphics*[width=8.5cm]{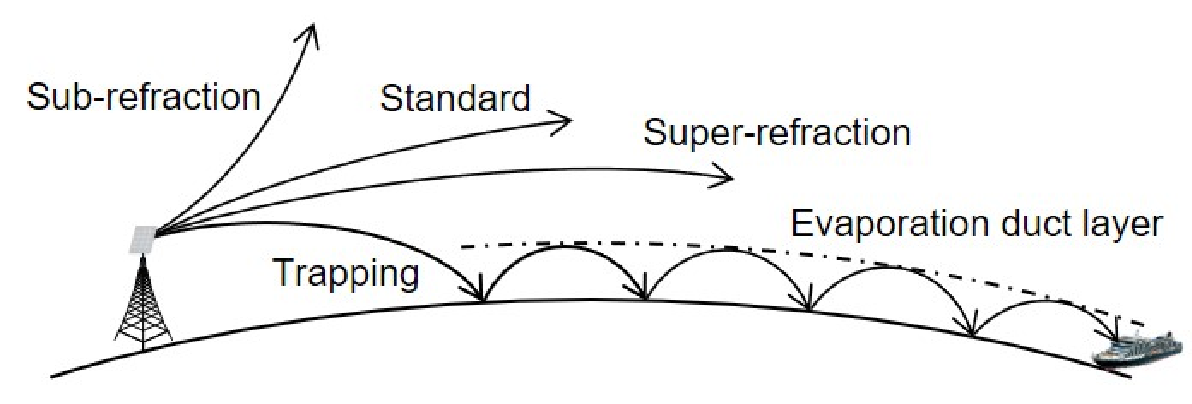} %这行为导入图片文件名称
\end{center}
\vspace*{-1mm}
\caption{Illustration of the four possible propagation paths due to atmospheric refraction and the evaporation duct.}\label{fig:2}  %这行为文章中显示图片的标题
%\vspace*{-2mm}
\end{figure}

\subsection{Improving Resource Utilization: Resource Management and Allocation Schemes}

The transmission efficiency of MCNs depends on the channel environment.
Therefore, advanced resource allocation schemes, such as dynamic beamforming and user scheduling techniques, can be used to take advantage of the dynamic changes of maritime channels.
For random and rapidly changing wireless channels, traditional resource allocation and utilization methods based on service statistics and characteristics are inefficient, as they lead to a significant decrease in the overall performance of the network. To deal with the dynamic changes in maritime channels from sea surface and weather conditions, it is necessary to fully exploit the characteristics of the MCN.

%The characterization and use of this feature involves electromagnetic propagation environment models and tomographic channel models. Based on the tomographic channel model, the use of large-scale channel information to explore a multi-point intelligent collaborative transmission optimization solution for business processes is a feasible technical approach. It can provide a new idea for the system to handle complex interference, efficiently allocate resources, and increase system capacity.

%The research on the optimal allocation of resources for maritime broadband communication systems is still in its infancy, and there are few related references.
The authors in \cite{p2005} used vertically spaced multiple antennas at the receiver side and proposed a frequency and time synchronization and scheduling scheme to overcome deep fading, assuming the two-path characteristic of maritime channels.
The authors in \cite{p359} proposed a service-oriented framework for the management of MCNs and developed three policy-based routing schemes using the framework.
In addition to the above works, WiMAX and delay-tolerant networking (DTN) technologies have been widely discussed for maritime communications \cite{p354}.
%The authors in  and WiMAX technologies for practical maritime communications.
The authors in \cite{p351} used the WiMAX-based mesh technology for ship-to-ship communications with DTN features and compared the performances of different routing schemes.
The authors in \cite{p356} studied the scheduling of data traffic tasks to optimize the network throughput and energy sustainability.
In \cite{p6022}, the authors proposed a joint backhaul and access link resource management scheme for the maritime mesh network to maximize the network capacity.

%, in practice, use of the storage-carrying-forward way to support intermittent communication, large delay,

%In addition to the above works discussing the resource management methods of MCNs,

%In practical applications, appropriate integration should be adopted according to different needs of the Internet.
%Although there are many different integrated networks, the development of more efficient integrated marine communication networks is still one of the research priorities in the future.

In contrast to terrestrial networks, user behaviour characteristics are useful in MCNs to improve its transmission efficiency,
%For the maritime environment,
since most marine users, such as passenger ships and cargo vessels, follow specific shipping lanes \cite{p1515}--\cite{p1518}.
The authors in \cite{p352} derived a model of ship encounter probability and used the model to analyse the data delivery ratio.
The authors in \cite{p353} proposed an architecture of delay-tolerant MCNs where the AIS is integrated to obtain the trip-related data of ships, and they optimized the routing performance utilizing ship contact opportunities.
%obtained ship movement data based on the existing AIS system and predicted the ship-ship encounter model.
%In \cite{p502}, the authors proposed an opportunistic routing scheme for delay-tolerant MCNs based on lane intersecting opportunities.
In \cite{p503} and \cite{p355}, the authors proposed energy-and-content-aware scheduling algorithms for video uploading in MCNs based on the deterministic network topology and the ship route traces, respectively.
%These studies utilized the predictability and stability of user movement, but did not fully take advantage of the physical characteristics of maritime channels.
%As

The studies in Section III.A have suggested that maritime channels consist of only a few strong propagation paths due to the limited number of scatterers, making the large-scale channel fading more dominant.
Thus, it is promising to allocate resources using the large-scale CSI only, which can be conveniently acquired from the location information in the MCN \cite{p1517}.
Previous studies have explored the performance gain achieved by power allocation \cite{p2101} and user scheduling \cite{p2102} techniques using only large-scale CSI and suggested that coordinated transmission with large-scale CSI is effective in
practical MIMO or distributed MIMO systems to improve the spectral efficiency and energy efficiency \cite{p2103}.
%analysed the spectral efficiency with only large-scale CSI,
%based on the location information predicted from the shipping lane .
%Therefore, acquiring long-term large-scale CSIT based on the location information predicted from the shipping lane is more feasible than obtaining instantaneous full CSIT, and is considered as a promising way to improve the coverage performance for practical MCNs.
%In addition, dynamic beam forming methods such as phased array antennas are used to distribute users over a wide range of areas and utilize the characteristics that marine users are more sparse than terrestrial users. According to the requirements of different service service quality, they can form smart devices on demand. Coverage to achieve optimization of coverage strength and breadth, efficient use of load power, spectrum and other resources. At the same time, by developing antenna coding and its design analysis methods, the theoretical problems of dynamic beam generation and sidelobe suppression can be systematically solved to achieve dynamic beam tracking.

% Table generated by Excel2LaTeX from sheet 'Sheet2'

In Table IV, we summarize the technologies to enhance transmission efficiency for MCNs, as well as the unique characteristics of maritime communications that have been utilized. From the table, we can see that it is promising to utilize the unique features of MCNs in terms of electromagnetic propagation environment, service requirement and vessel movement for more efficient transmission.
Specifically, since maritime channels are susceptible to sea surface conditions and atmospheric conditions, future MCNs need to be able to perceive environmental information, such as the sea level, temperature, humidity, and wind speed, to make more accurate prediction of the CSI and then intelligently utilize the dynamic changes of maritime channels for more efficient transmission.

\begin{table*}[htbp]
  \centering
  \caption{Technologies to Enhance Transmission Efficiency for MCNs.}
  \scriptsize
    \begin{tabular}{|p{12em}|p{9em}|p{15em}|p{25em}|}
    \hline
    \multicolumn{1}{|c|}{\textbf{Goal}} & \textbf{Scheme} & \textbf{Characteristics of MCNs Used} & \multicolumn{1}{c|}{\textbf{Contributions}} \bigstrut\\
    \hline
    \multicolumn{1}{|c|}{\multirow{2}[4]{*}{reducing transmission loss}} & microwave scattering & evaporation duct over sea & setting up beyond-LOS maritime communications: 78-km \cite{p343},  100-km \cite{p344}\cite{p345} \bigstrut\\
\cline{2-4}    \multicolumn{1}{|c|}{} & \multicolumn{1}{p{9.625em}|}{frequency-time scheduling} & \multicolumn{1}{p{16.875em}|}{two-path maritime channels} & exploiting vertically spaced multiple antennas at the Rx. to overcome deep fading \cite{p2005} \bigstrut\\
    \hline
    \multicolumn{1}{|c|}{reducing transmission delay} & routing & \multicolumn{1}{p{16.875em}|}{delay-tolerant maritime services} & a service-oriented framework and three policy-based routing schemes \cite{p359}, DTN technologies \cite{p354}\cite{p351} \bigstrut\\
    \hline
    \multicolumn{1}{|c|}{\multirow{2}[4]{*}{improving network throughput}} & traffic scheduling & content-aware maritime applications & data traffic task scheduling \cite{p356} \bigstrut\\
\cline{2-4}    \multicolumn{1}{|c|}{} & resource allocation & heterogeneous networking & joint backhaul and access link resource management \cite{p6022} \bigstrut\\
    \hline
    improving data delivery ratio & routing & \multicolumn{1}{c|}{\multirow{2}[4]{*}{user behaviours (such as shipping lanes)}} & a model of ship encounter probability \cite{p352}, an architecture integrating the AIS  \cite{p353} \bigstrut\\
\cline{1-2}\cline{4-4}    improving energy efficiency & traffic scheduling &       & energy-and-content-aware scheduling algorithms for video uploading based on the deterministic network topology \cite{p503} and the ship route traces \cite{p355} \bigstrut\\
    \hline
    \end{tabular}%
  \label{tab:addlabel}%
\end{table*}%

\section{Increasing Broadband Coverage}\label{sec:2}

%\subsection{Limited number of BS sites}

%%with millions of people travelling all around the World in Ships and Ferries.
%%Satellites can cover a wide area over sea.
%Satellite communications systems can cover a wide area over sea, but most of them only provide narrowband services.
%%For example, the communication rates of Iridium and Globalstar are only 4.8 kbps and 9 kbps, respectively \cite{p513}\cite{p514}.
%The newly launched high-throughput satellites, such as EchoStar-19 and Inmarsat-5, have enabled broadband maritime coverage.
%However, the cost of ship-borne equipment is still very high \cite{p1501}.
%%For example, the cost of installing ship-borne equipment for Inmarsat (Fleet 77) is approximately \$28000, including the antennas, terminal, handset, power supply, etc. \cite{p1501}.
%Data from the AIS show that there are nearly 80,000 ships sailing simultaneously around the world, less than 25,000 of which are high-end ships (with a load of more than 10,000 tons) that may afford the ship-borne equipment for high-throughput satellite communications.

%, and can provide low-speed or high-speed data services depending the narrowband or broadband access.
%However, satellite-based communications are greatly affected by climatic conditions and the marine environment, resulting in low reliability.
%As represented by comets and global stars \cite{p513}, the low-orbit communication satellites adopt a globally uniform coverage, but the communication rates are 4.8 kbps and 9 kbps, respectively \cite{p514}.
\textcolor{black}{The previous section has focused on enhancing the transmission efficiency for the MCN from the link-level perspective. However, due to the quite limited BS sites in an MCN, only link-level enhancement is not enough to ensure seamless wide-area coverage. For this reason, we focus on the utilization and coordination of all available wireless coverage approaches, and introduce the key technologies for increasing broadband coverage from a system/network-level perspective in this section.}

In addition to maritime satellites, shore \& island-based BSs can be built to extend the coverage of terrestrial networks to the ocean. UAVs, high-end ships and offshore lighthouses can be exploited as well to serve as maritime BSs.
The coverage performance of the MCN depends largely on the abovementioned geometrically available BS sites.
%In a terrestrial cellular network, it is possible to extend the coverage by installing more BSs.
%However, in an MCN, the available BS sites are very limited.
%On the other hand,
Due to the limited onshore BS sites and the high mobility of the ship-borne BSs, aerial BSs and users, the topology of the MCN is highly irregular.
There always exist blind zones within the coverage area.
%However,
When the BS increases its transmission power, it will also generate strong co-channel interference to users served by the neighbouring BSs.

Considering these problems, the MCN has to make full use of the available BSs, including onshore BSs, ship-borne BSs, aerial BSs and satellites, as depicted in Figure 6. By using multi-hop wireless networking and satellite-terrestrial integration technologies, the BSs can work cooperatively to increase broadband coverage.
%Specifically, multi-hop wireless networking, satellite-terrestrial integration, phased-array antennas, and microwave scattering techniques are utilized.
Advanced transmission techniques, such as dynamic beamforming and microwave scattering, can also be used to reduce the signal attenuation and extend the coverage of a single BS.
%In addition, satellite-terrestrial coordination techniques need to be studied to mitigate the complex co-channel interference.

%specifically, due to the limited spectrum resources, some systems and

\begin{figure} [t]
\vspace*{1mm}
\begin{center}
\includegraphics*[width=9cm]{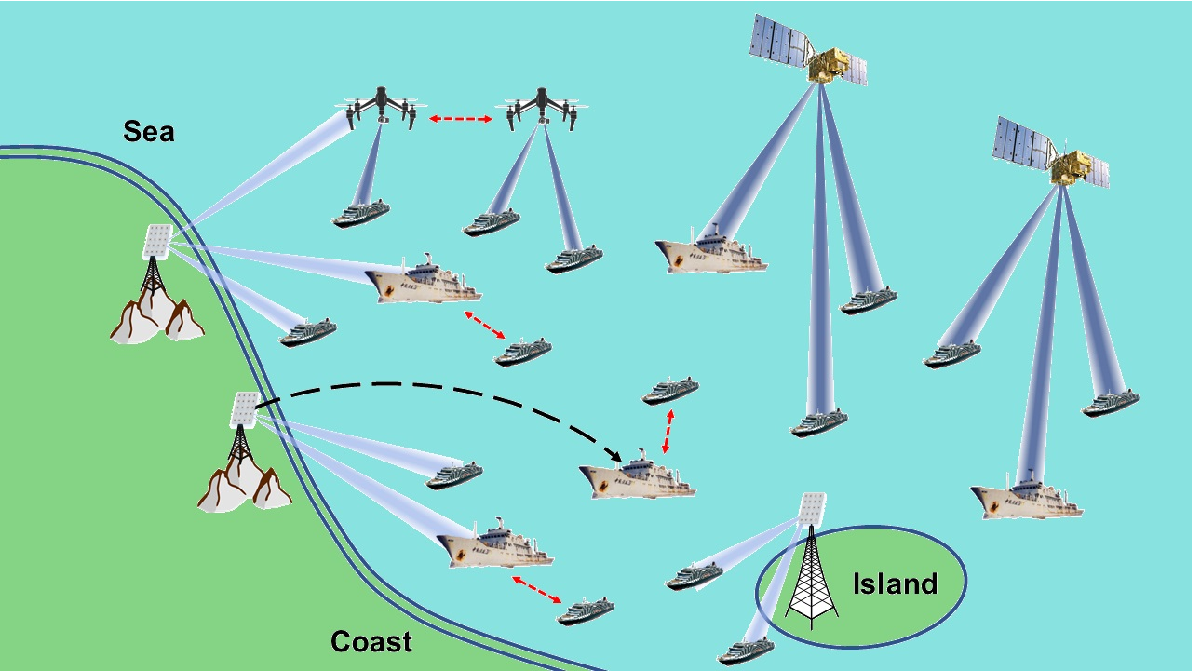} %这行为导入图片文件名称
\end{center}
%\vspace*{-4mm}
\caption{Exploiting onshore BSs, ship-borne BSs, aerial BSs and satellites to extend the maritime coverage.}\label{fig:2}  %这行为文章中显示图片的标题
%\vspace*{-2mm}
\end{figure}

\subsection{Building and Exploiting Offshore BSs: Multi-hop Networking of Ship-borne and UAV-enabled BSs}

To achieve wider coverage for MCNs, the authors in \cite{p901} and \cite{p910} proposed the ad hoc networks Maritime-MANET and Nautical Ad hoc Network (NANET), respectively.
Similarly, the authors in \cite{p351} proposed a WiMAX-based mesh network to provide delay-tolerant maritime communication services.
%, and compared the performance of different routing schemes.
To improve the efficiency of ship-to-ship communications in these MCNs, \cite{p901} used multiple directional antennas, \cite{p1003} used virtual MIMO technologies, and \cite{p512} and \cite{p6024} used
two relaying schemes. Additionally, in \cite{p914}, a novel handover protocol was proposed. In \cite{p915}, a distributed adaptive time slot allocation scheme was proposed, while in \cite{p916}, a cognition-enhanced mesh medium access control (MAC) protocol was proposed.
Further, the authors in \cite{p511} proposed an integrated MCN consisting of NANETs, terrestrial cellular networks, and satellite networks, in order to meet the requirements of various services.

%In order to systematically describe the entire maritime communication framework, \cite{p511} proposed an integrated MCN consisting of a VANET, a cellular network, and a satellite network.
%%In this system, all similar ships first access the VANET and then connect to the cellular network, and only communicate with the satellite if the cellular network fails.
%
%
%At the same time,  to provide maritime users with a variety of services such as positioning, navigation, distress, voice, weather forecasting, etc., by comprehensively utilizing VANETs, cellular networks and satellite networks \cite{p910}--\cite{p911}.

%\begin{figure} [htb]
%\begin{center}
%\includegraphics*[width=12cm]{mesh.eps} %这行为导入图片文件名称
%\end{center}
%\vspace*{-4mm} \caption{Architecture of a multi-hop wireless mesh network for maritime communications proposed in \cite{p901}.}\label{fig:2}  %这行为文章中显示图片的标题
%\vspace*{-2mm}
%\end{figure}

Various routing methods and protocols have been proposed for terrestrial delay-tolerant ad hoc networks \cite{p904}\cite{p1108}, such as epidemic routing \cite{p903}, probabilistic routing \cite{p905}, spray and wait \cite{p906}, network coding schemes \cite{p907}\cite{p908}, Optimized Link State Routing (OLSR) \cite{p917}, Ad Hoc On Demand Distance Vector (AODV) Routing \cite{p918}--\cite{p1023}, and Ad Hoc On Demand Multipath Distance Vector (AOMDV) \cite{p919}.
However, these schemes have poor performances in maritime communications, due to the large delivery delay and
low delivery ratio from the lower user density \cite{p909}. Therefore, routing protocols custom-made for maritime mesh networks are required \cite{p920}--\cite{p921}.

%In contrast to terrestrial networks, user behavior characteristics are exploitable in MCNs, since most marine users such as passenger ships and cargo vessels follow specific shipping lanes.
In \cite{p502}, the authors proposed an opportunistic routing scheme for delay-tolerant MCNs based on lane intersecting opportunities.
In \cite{p503}, the authors proposed three offline scheduling algorithms for video uploading in MCNs based on the deterministic network topology.
In \cite{p2006}, the authors proposed a route maintenance method for maritime sensor networks based on ring broadcast mechanism.
\textcolor{black}{In addition, the authors of \cite{p20201206-2} and \cite{p20201206-3} proposed two secure and efficient routing protocols for the Internet of Mobile Things based on movement prediction.}
These studies utilized the predictability and stability of user movement but did not take full advantage of the physical characteristics of maritime channels.
Different features and applicable scenarios of representative routing protocols for maritime communications are listed in Table V.
Note that the height and angle of the ship-borne antenna are rapidly changing due to the fluctuation of the sea surface. In addition, maritime channel fading is sensitive to antenna height and angle \cite{p913}. To solve this problem, we need to establish a sensitivity model for the received signal strength, the height of the ship-borne antennas, and the sea surface fluctuation intensity, based on which we can optimize the routing algorithm in MCNs to reduce the packet loss rate and network delay.

%\begin{figure} [htb]
%\begin{center}
%\includegraphics*[width=15cm]{lane-based-routing.eps} %这行为导入图片文件名称
%\end{center}
%\vspace*{-4mm} \caption{Lane-Based Optimal Routing Protocol for Delay-Tolerant Maritime Networks in \cite{p502}.}\label{fig:2}  %这行为文章中显示图片的标题
%\vspace*{-2mm}
%\end{figure}

\begin{table*}
\centering
\caption{Routing protocols for maritime communications.}
\scriptsize
\begin{tabular}{|c|c|c|c|}  %通过添加|来表示是否需要绘制竖线
\hline  % 在表格最上方绘制横线
\textbf{Reference} & \textbf{Protocol}  & \textbf{Feature} & \textbf{Applicable Scenario}\\
\hline  %在第一行和第二行之间绘制横线
\cite{p904} &  Routing Application for Parallel Computation of Discharge (RAPID) & Replica-based (flooding) & Ships in low density\\
\hline  %在第一行和第二行之间绘制横线
\cite{p903} &  Epidemic Routing (ER) & Replica-based (flooding) & Ships in low density\\
\hline  %在第一行和第二行之间绘制横线
\cite{p905} & Probabilistic Routing (PR) & Replica-based (flooding) & Ships in low density\\
\hline  %在第一行和第二行之间绘制横线
\cite{p906} & Spray and Wait (SaW) & Replica-based (flooding) & Ships in low density\\
\hline  %在第一行和第二行之间绘制横线
\cite{p907} & Estimation Based Erasure Coding (EBEC) & Coding based & Ships in low density\\
\hline  %在第一行和第二行之间绘制横线
\cite{p908} & Hybrid Erasure coding (HEC) & Coding based & Ships in low density\\
\hline  %在第一行和第二行之间绘制横线
\cite{p917} & Optimized Link State Routing (OLSR) & Regular & Ships in good density\\
\hline  %在第一行和第二行之间绘制横线
\cite{p918} & Ad Hoc On Demand Distance Vector (AODV) & Regular & Ships in good density\\
\hline  %在第一行和第二行之间绘制横线
\cite{p919} & Ad Hoc On Demand Multi-path Distance Vector (AOMDV) & Regular & Ships in good density\\
\hline  %在第一行和第二行之间绘制横线
\cite{p502} & Lane-Based Opportunistic Routing (LanePost) & Knowledge based & Ships in low density\\
\hline  %在第一行和第二行之间绘制横线
\cite{p922} & Geographical Routing (GR) & Knowledge based & Ships in low density\\
\hline  %在第一行和第二行之间绘制横线
\cite{p923} & Gradient Routing Based on Link Metrics (GR-LM) & Knowledge based & Ships in low density\\
\hline  %在第一行和第二行之间绘制横线
\end{tabular}
\end{table*}

%\begin{figure} [htb]
%\begin{center}
%\includegraphics*[width=12cm]{sensitivity.eps} %这行为导入图片文件名称
%\end{center}
%\vspace*{-4mm} \caption{Sensitivity of 2-ray maritime channel model under sea-level fluctuations.}\label{fig:2}  %这行为文章中显示图片的标题
%\vspace*{-2mm}
%\end{figure}

Moreover, marine traffic also fluctuates over time, resulting in changes in BS loading. Using BS switching, the MCN can shut down some low-loaded BSs when the traffic is low to not only support current users but also save energy and reduce interference to neighbouring users. At present, the switch selection methods applicable to terrestrial fixed BSs are based on BS performance indicators such as coverage, cell load, and neighbouring cell interference, to provide a BS deployment scheme and a switching method. Unlike terrestrial BSs, a ship-borne BS has the following two characteristics: First, the power resources are limited, so it is more important to save energy. Second, the on-board BS has high mobility. If the switch selection method for terrestrial fixed BS is applied to maritime communications, either the BS switch operation will be too frequent or the BS switch configuration for a period of time will not meet the user needs. Therefore, existing BS switch selection methods do not apply to the onboard BS. Switch selection methods for ship-borne BSs in maritime communications need to be investigated. For example, the authors in \cite{p1016} proposed a ship-borne BS sleeping control and power allocation scheme for the MCN based on the sailing route to enhance the robustness of dynamic coverage.

\textcolor{black}{UAVs are believed to be useful and efficient for promoting connectivity in vehicular ad hoc networks \cite{p20201206-4}.}
In addition to ship-borne BSs, UAVs can be exploited to serve as aerial maritime BSs, or relay nodes to extend the coverage of the MCN.
Specifically, the coordination among UAVs can provide a multi-hop network, such as a mesh network, where the flight trajectory, routing strategy and transmission method are optimized.
In \cite{p2104}, the authors focused on the reliability, and optimized the altitude of the UAV as a relaying station.
In \cite{p2105}, the authors considered UAV-aided data collection for the maritime IoT and optimized the transmit power and duration of all devices to maximize the data collection efficiency. Due to their agile manoeuvrability, UAVs are considered effective tools to achieve dynamic and flexible coverage for MCN.
Despite that, utilizing UAVs for maritime IoT applications (such as data gathering) still faces challenges. For example, it is difficult to recharge battery-equipped maritime IoT devices, so the energy constraint must be considered to optimize the communication strategy. In addition, it is difficult to acquire perfect CSI (including the random small-scale CSI) due to the sea wave movement.

\subsection{Utilizing High-throughput Satellites: Multi-spot Beams and Satellite-terrestrial Cooperation}

In addition to building a mesh network using ship-borne BSs and UAVs, satellites can also be exploited to extend the coverage of MCNs.
The utilization of satellite communications for maritime coverage has been widely reported in the literature \cite{p705}--\cite{p1011}.
%In \cite{p701}, the authors considered the feasibility of utilizing illuminating signals sent by Inmarsat for maritime surveillance and navigation, especially for marine obstacle avoidance. In \cite{p717}, the authors introduced the utilization of satellite-based AIS receivers to extend traffic monitoring zones to open seas, as well as the challenges of message collision in high traffic zones. More recently, the authors in \cite{p702} proposed a parallel signal processing architecture and algorithms for satellite-based AIS to cope with the message collision in dense maritime zones, and reduce the downlink power, bandwidth, and latency.
Although satellite communications have a wide coverage, they are limited by their high latency and low data rate.
To enhance broadband coverage of maritime satellites, spot beam and frequency reuse technologies have been studied.
Since a narrower beam width leads to a higher antenna gain, the use of spot beam technologies can increase the spectral efficiency, and allow maritime users to use smaller satellite terminals \cite{p705}--\cite{p712}.
Further, the use of multi-spot beams allows beams that are far apart to reuse frequency.
Frequency reuse is an effective way to improve spectral efficiency, but it may generate strong inter-beam interference due to the non-zero side lobes \cite{p707}.
Therefore, side lobe suppression technologies are required for the use of multi-spot beams, and there is a trade-off between the number of spot beams and the distance between frequency-reuse beams \cite{p720}\cite{p719}.
It should be noted that, in maritime communications, the density of vessels/platforms/islands is low, while the users are clustered thereon. Therefore, using multi-spot beams is an effective way to enhance broadband coverage for MCNs \cite{p721}\cite{p722}.

Since terrestrial networks in general have high capacity but limited coverage, while satellites have wide coverage but a low data rate, an integrated satellite-terrestrial network (ISTN) is a promising way to enable seamless broadband coverage, taking advantage of both networks \cite{p2106}.
%Up to now, a number of papers have reviewed ISTNs from different perspectives.
%%\cite{7245586} investigated the channel models and the terrestrial interference for satellite television broadcast.
%For example, \cite{p3001} presented a review of several important issues for ISTNs, such as network design and optimization.
%Reference \cite{4455067} pointed out some research directions for ISTNs, focusing on the network layer and the transport layer.
%Reference \cite{7463014} made a survey on QoS performance for the ISTN.
%Reference \cite{1355878} described an ISTN for multimedia broadcasting services and the required resource management strategies.
%%\cite{1031995} and \cite{5714025} investigated the challenges, opportunities, and solutions for HSTNs.
%Reference \cite{p5206} investigated the problem of cooperative transmission in future ISTNs.
%Reference \cite{5683428} presented the concept and key issues of cognitive ISTNs.
%Further, the authors in \cite{p1901} considered a cognitive ISTN where the terrestrial system shares resources with the satellite network with constraint on the interference temperature and compared the outage performances of different secondary transmission schemes.
%In \cite{p1902}, the authors considered an ISTN where cognitive relay stations are set to forward the received signals from satellites, and proposed a power allocation scheme to maximize the achievable rate.
%In \cite{p2107}, the authors considered the integration of UAVs into the satellite network and proposed a power allocation algorithm to improve user fairness.
Specifically, for maritime communications, the authors in \cite{p715} proposed intelligent middleware and link-specific protocols for the coordination of maritime mesh networks and satellite communication networks, as depicted in Figure 7. The authors in \cite{p704} considered a hybrid Satellite-MANET consisting of GEO, MEO, and LEO satellites and terrestrial MANET. They analysed the distribution of coverage radius
for full connection and proposed a multi-hop routing protocol to minimize the end-to-end delay.
The authors in \cite{p1011} proposed an OceanNet Backhaul Link Selection (OBLS) algorithm  to select the optimal backhaul links with the best signal to noise ratio (SNR). Further, the authors implemented the proposed algorithm in a hardware test-bed.

\begin{figure} [t]
%\vspace*{1mm}
\begin{center}
\includegraphics*[width=9cm]{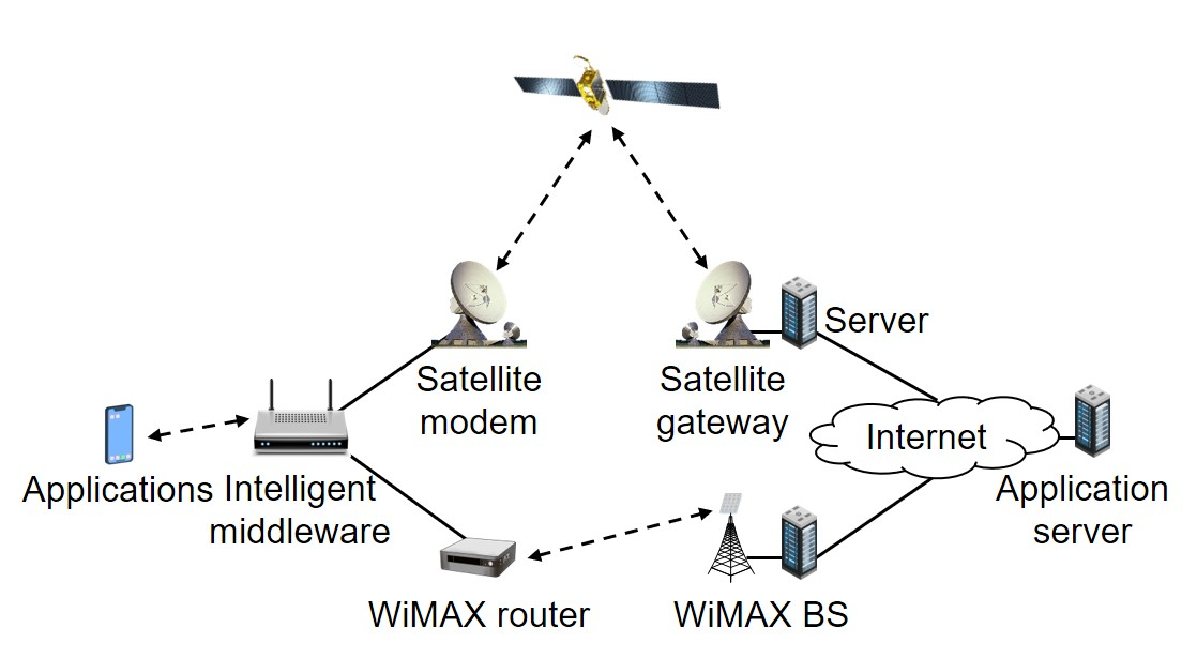} %这行为导入图片文件名称
\end{center}
\vspace*{-1mm}
\caption{Using intelligent middleware for satellite-terrestrial cooperation \cite{p715}.}\label{fig:2}  %这行为文章中显示图片的标题
%\vspace*{-2mm}
\end{figure}

It should be noted that the round-trip time in satellite communications is much longer than cellular communications. This will greatly degrade the quality of service (QoS) and quality of experience (QoE), especially for real-time video needed in marine rescue.
To tackle this problem, the authors in \cite{7962768} proposed a backhaul activation scheme to minimize the traffic delivery time for a multi-hop ISTN.
In addition, caching strategies have been adopted to reduce the accumulative delay. In \cite{p20228}, the authors proposed a back-tracing partition based on-path caching algorithm for ISTNs to reduce the overheads and access delay. In \cite{p20238}, the authors proposed a QoE-driven caching placement scheme for video streaming in the ISTN, considering the social relationship among users. In \cite{p20250}, the authors presented a secure hybrid in-network caching scheme for multimedia content streaming in the ISTN.

It is also possible to increase the broadband coverage for MCNs by building large-scale LEO satellite constellations.
SpaceX plans to launch approximately 12,000 Starlink LEO satellites. There will be two Starlink satellite constellations: one containing 4,409 satellites and the other containing 7,518 satellites.
%The second, slightly larger group of satellites operated at a slightly lower altitude.
%Both will provide affordable Internet services.
SpaceX plans to provide affordable Internet services with delays between 25 ms and 35 ms, which will make its services comparable to cable and fiber optics.
In addition to SpaceX, many companies such as OneWeb, TeleSat, and Amazon hope to provide Internet access services to more people by deploying small LEO satellite networks \cite{p10404}--\cite{p10407}.
There are many challenges for large-scale LEO satellite constellations, such as mobility management, resource allocation, and security. For example, conventional Internet protocols have large signalling overhead and handover delay due to the frequent changes of LEO satellites' point of attachment (PoA) in the ISTN, and methods of installing mobility logic in the software defined network (SDN) controller to address the PoA variation need to be studied.

\subsection{Reducing Signal Attenuation: Phased-array Antennas and Beam Scheduling Techniques}

Directional beams are commonly used to widen the coverage area. The concept of beamforming was introduced in 5G, where the beamforming vectors are calculated based on the MIMO CSI \cite{p1604}.
%However, in the architecture of 5G, the calculation of channel is complex with expensive hardware cost, and the frequency of 5G is high and causes a higher path loss.
%As a better solution, 4G LTE (Long-term Evolution) adopts OFDMA (Orthogonal Frequency Division Multiple Access) as a physical layer modulation method to improve spectral efficiency, and uses a variety of algorithms to increase the data rate, which can help achieve a downlink peak rate of 100 Mbps. Moreover, with the development of LTE in recent years, the cost of hardware is greatly reduced.
%The beam forming in 4G is digital, and the hardware cost of digital beamforming is expensive.
In the scenario of maritime communications, the density of vessels is low and the users are clustered in a small area (ship/platform),
%the channel is close to direct path [3],
which makes it easy to determine the beam directions according to the users' geographical location. Thus, it is convenient for MCNs to use phased array antennas with analogue beamforming to reduce the cost.
The use of phased array antennas can effectively increase the coverage of MCNs with a lower cost based on the existing LTE networks \cite{p1605}, as the directions of antennas can be determined according to the location of users.

The user location information can be obtained from the AIS.
%The user location can be accessed by the AIS.
The BS receives the location information and steers the antennas to point to the selected directions \cite{p1606}.
%It is noticed that broadcast data with lower data rate is transmitted through an omnidirectional antenna.
Note that directional beams can point to a narrower range of directions than omni-directional beams to decrease the signal attenuation, and the beams need to be dynamically scheduled, for higher throughput or wider coverage \cite{p1607}--\cite{p1701}.
%It is assumed that the beams are discrete and orthogonal to each other, as they obtain higher computational efficiency, and reduce the number of bits required for the transmission of position data \cite{p1203}. In the meantime, phased array antennas are supposed to fix to one direction in one time in order to have a large power gain \cite{p1701}.

%\begin{figure} [htb]
%\begin{center}
%\includegraphics*[width=9cm]{BEAM.eps} %这行为导入图片文件名称
%\end{center}
%\vspace*{-4mm} \caption{Dynamic beam scheduling using phased array antennas.}\label{fig:2}  %这行为文章中显示图片的标题
%\vspace*{-2mm}
%\end{figure}

\subsection{Exploiting Microwave Scattering for Over-the-horizon Coverage of Islands/Platforms}

The earth's atmosphere can be divided into the ionosphere, the stratosphere, and the troposphere.
The troposphere is the atmosphere from earth surface to an average altitude of 10--12 km.
The turbulence and the inhomogeneous medium in the troposphere can scatter incident microwaves to allow over-the-horizon communications.
Microwave scattering communications have the advantages of long distance, large capacity, high security, and high flexibility. Therefore, microwave scattering is very suitable for providing communication services for users in environmentally harsh areas, such as mountains, deserts, and oceans \cite{p1014}--\cite{p1017}.

%Microwave scattering is a method to achieve over-the-horizon communications.
%It utilizes the inhomogeneous atmospheric components in the troposphere to generate forward-scattering of microwave signals to achieve over-the-horizon propagation.
%The troposphere in the atmosphere has a large number of constantly changing turbulence.
%The turbulence scatters the waves around.
%When the wavelength equals to the size of the turbulence, the main radiation direction is in the front, and some of the energy is turned to the ground.
%Microwave scattering works as follows: The transmitter emits microwave signals to the troposphere at a certain elevation angle, and then a large number of signal components are scattered forward by the scatterers (also a small amount of refraction and reflection components) to the receiver to complete one-hop over-the-horizon propagation.
%The microwave scattering single-hop span is about tens to hundreds of kilometers, the transmission rate can reach more than 34 Mbps, and the transmission delay is several milliseconds; the channel is not affected by the harsh natural environment such as lightning, aurora, magnetic storm and sunspot.

%The authors made a comprehensive and systematic study and summary of microwave scattering from the perspectives of communication mechanism, transmission loss, fading characteristics and statistical characteristics of each feature.

%At present, microwave scattering is mainly used for long-distance communication on land, but

%While the number of scatterers on the sea surface is limited, there are more

The number of scatterers in the troposphere over the oceans is much larger than that in the troposphere over the ground, due to more frequent atmospheric flows. Thus, the transmission distance using microwave scattering in maritime communications is larger than that in terrestrial communications \cite{p1206}\cite{p1001}. Until now, many experimental links have been set up for over-the-horizon maritime communications using microwave frequency band, such as the 5.8 GHz band \cite{p1014}, and the 2.2 GHz band \cite{p1017}.

High-power microwave antennas or large-scale antenna arrays often require compensation for the transmission loss. Therefore, microwave scattering communications may not be cost-effective and are mainly used for the coverage of islands, warships, and drilling platforms \cite{p1002}.

%which makes the forward scatter signal more concentrated. In theory, the ocean microwave scattering communication distance is longer and the capacity is larger. The performance of maritime communications using microwave scattering is much better than that of terrestrial communications.
%Moreover, compared with other maritime communication methods, the channel environment of scattering communications is better than MF/HF wireless communications, its transmission distance is greater than VHF wireless communications, and its operating cost is lower than satellite communications.
%The scatter communication equipment is flexible and can be deployed either as a fixed station (island or floating platform) or as a mobile station (onboard).
%Therefore, microwave scattering is more suitable for long-distance (hundreds of kilometers) maritime communications, especially for islands and offshore platforms .

% Table generated by Excel2LaTeX from sheet 'Sheet4'
\begin{table*}[htbp]
  \centering
\caption{Technologies to Increase Broadband Coverage for MCNs.}
  \scriptsize
    \begin{tabular}{|l|l|l|p{25.19em}|}
    \hline
    \multicolumn{1}{|p{4.875em}|}{\textbf{Approach}} & \multicolumn{1}{p{11.815em}|}{\textbf{Goal}} & \multicolumn{1}{p{4.25em}|}{\textbf{Scheme}} & \textbf{Technology} \bigstrut\\
    \hline
    \multicolumn{1}{|l|}{\multirow{11}[22]{*}{exploiting ship-borne BSs}} & \multicolumn{1}{l|}{\multirow{6}[12]{*}{\shortstack{providing delay-tolerant maritime \\communication services}}} & \multicolumn{1}{l|}{\multirow{3}[6]{*}{network architecture}} & WiMAX-based mesh network \cite{p351} \bigstrut\\
\cline{4-4}          &       &       & Maritime MANET \cite{p901} \bigstrut\\
\cline{4-4}          &       &       & Nautical ad hoc network \cite{p910} \bigstrut\\
\cline{3-4}          &       & \multicolumn{1}{l|}{\multirow{3}[6]{*}{resource allocation}} & dynamic subnet relay \cite{p512} \bigstrut\\
\cline{4-4}          &       &       & energy-efficient handover \cite{p914} \bigstrut\\
\cline{4-4}          &       &       & routing (shown in Table V) \bigstrut\\
\cline{2-4}          & \multicolumn{1}{l|}{\multirow{5}[10]{*}{\shortstack{improving the efficiency of \\ship-to-ship communications}}} & \multicolumn{1}{l|}{\multirow{2}[4]{*}{multiple antennas}} & multiple directional antennas \cite{p901} \bigstrut\\
\cline{4-4}          &       &       & virtual MIMO technologies \cite{p1003} \bigstrut\\
\cline{3-4}          &       & \multicolumn{1}{l|}{\multirow{3}[6]{*}{resource allocation}} & distributed adaptive time slot allocation \cite{p915},   \bigstrut\\
\cline{4-4}          &       &       & cognition-enhanced mesh MAC protocol \cite{p916} \bigstrut\\
\cline{4-4}          &       &       & ship-borne BS sleeping control and power allocation \cite{p1016} \bigstrut\\
    \hline
    \multicolumn{1}{|l|}{\multirow{5}[10]{*}{utilizing high-throughput satellites}} & \multicolumn{1}{l|}{\multirow{2}[4]{*}{\shortstack{improving the throughput of \\satellite-to-ship communications}}} & \multicolumn{1}{l|}{\multirow{2}[4]{*}{smart antennas}} & smart satellite terminals \cite{p706}\cite{p712} \bigstrut\\
\cline{4-4}          &       &       & phased array antennas \cite{p719}--\cite{p722} \bigstrut\\
\cline{2-4}          & \multicolumn{1}{l|}{\multirow{3}[6]{*}{satellite-MANET coordination}} & \multicolumn{1}{l|}{\multirow{3}[6]{*}{network protocol}} & intelligent middleware and link-specific protocol \cite{p715} \bigstrut\\
\cline{4-4}          &       &       & multi-hop routing protocol \cite{p704} \bigstrut\\
\cline{4-4}          &       &       & backhaul link selection algorithm \cite{p1011} \bigstrut\\
    \hline
    \multicolumn{1}{|l|}{\multirow{4}[8]{*}{extending the coverage of a single BS}} & \multicolumn{1}{l|}{\multirow{2}[4]{*}{reducing  propagation loss}} & \multicolumn{1}{l|}{\multirow{2}[4]{*}{directional beams}} & MIMO transmit diversity and multiplexing \cite{p1203} \bigstrut\\
\cline{4-4}          &       &       & location-aware dynamic beam scheduling \cite{p1701} \bigstrut\\
\cline{2-4}          & \multicolumn{1}{l|}{\multirow{2}[4]{*}{\shortstack{achieving over-the-horizon \\communications}}} & \multicolumn{1}{l|}{\multirow{2}[4]{*}{microwave scattering}} & experimental microwave links \cite{p1015} \bigstrut\\
\cline{4-4}          &       &       & novel lightweight antennas \cite{p1017} \bigstrut\\
    \hline
    \multicolumn{1}{|l|}{\multirow{5}[10]{*}{interference alleviation}} & \multicolumn{1}{l|}{\multirow{3}[6]{*}{interference analysis}} & \multicolumn{1}{p{4.25em}|}{interference modelling} & modelling of interferences on other satellites in maritime satellite communications \cite{p707} \bigstrut\\
\cline{3-4}          &       & \multicolumn{1}{l|}{\multirow{2}[4]{*}{interference simulation}} & analysis and simulation of interference produced to the fixed service receivers by the mobile satellite service \cite{p720} \bigstrut\\
\cline{4-4}          &       &       & analysing the co-channel interference from maritime mobile earth station to 5G mobile service  \cite{p1006}\cite{p6011} \bigstrut\\
\cline{2-4}          & \multicolumn{1}{l|}{\multirow{2}[4]{*}{interference coordination}} & \multicolumn{1}{l|}{\multirow{2}[4]{*}{resource allocation}} & pilot scheduling and power allocation \cite{p1018} \bigstrut\\
\cline{4-4}          &       &       & radio resource block allocation \cite{p1010} \bigstrut\\
    \hline
    \end{tabular}%
  \label{tab:addlabel}%
\end{table*}%

\subsection{Interference Alleviation for Irregular Network Topology}

Due to the limited spectrum resources, some systems and beams of the MCN have to be frequency-multiplexed, and the interference model is complicated. Traditional methods often deal with the inter-system interference and intra-system interference independently. However, the MCN often has an irregular topology, and its coupling between the systems and the intra-system interference is very strong \cite{p1007}--\cite{p1009}.
To solve this problem, one can either schedule the beam resources between the satellite and the terrestrial network, study the optimal beam design method, or suppress inter-beam interference.
In \cite{p10090725}, an antenna selection algorithm was proposed for hierarchical maritime radio networks towards better coverage and cost-effectiveness.
%At the same time, in order to reduce the complexity, the centralized processing is converted to distributed processing, based on the probability map model and message passing algorithms to explore low-complex multi-user joint precoding techniques.
In \cite{p707}, the authors investigated the impact of diffracted waves from the structures of a ship on the received signal levels and found a relationship between the clearance angles and interference.
In \cite{p720}, the authors focused on the interference between the mobile satellite service and the fixed service in a maritime environment in the Ku band, and analysed the mobile earth station's transmit power, antenna gain, speed, and the propagation environment.
More recently, the authors in \cite{p1006} and \cite{p6011} analysed the co-channel interference from maritime mobile earth station to 5G mobile service and suggested that a separation distance should be set to guarantee the QoS for 5G outdoor environments.

Due to the long distance required to cover the sea surface, the propagation path needs to take into account the influence of the curvature of the earth.
For sea surface coverage, the wireless signal travels very far due to the small loss of radio wave transmission.
At this point, the influence of the curvature of the earth on the sea surface must be considered, and the sea surface cannot be regarded as a plane.
Therefore, the height of the antenna is directly related to the coverage distance \cite{p1005}\cite{p1207}. If the antenna height is too low, it will reduce the coverage of the BS. If the antenna height is too high, it will cause pilot contamination between neighbouring cells. Therefore, the antenna height must be carefully adjusted for maritime communications.

%The MCN has limited BS sites, and usually meets the coverage requirements of different BSs by changing the antenna configuration of the BS, and the baseband controls the antenna beam to serve multiple users through signal processing. To this end, the beam design relies on CSI between the transmitting and receiving antennas, but the CSI acquired from the baseband side is coupled to the RF parameters such as beam pointing and width. In response to this problem, it is possible to use the channel correlation between users on the baseband side to group users and design the direction and width of the beam, based on the user's spatial-temporal distribution. In addition, the coupling relationship between channel state and RF parameters should be studied, in order to propose low complexity iterative optimization method which further improves the resource efficiency of the network.

%Besides,
On the other hand, when the BS covers the remote users with high power, it will generate strong interference to users served by the neighbouring BSs, causing the near-far effect. The removal of interference requires the CSI, but the pilot transmitted by the nearby users also generates strong interference to the pilot transmitted by the remote user, resulting in inaccurate channel estimation. Thus, the pilots need to be carefully designed \cite{p1018}.

Up to now, the key technologies to extend the coverage of MCNs, such as multi-hop wireless networking of ship-borne BSs and UAVs, satellite-terrestrial cooperation, dynamic beam scheduling, microwave scattering, and interference management, have been studied. We summarize the technologies in Table VI.
To give full play to the advantages of the abovementioned methods, a heterogeneous network can be formed by coordinating maritime satellites, shore \& island-based BSs, ship-borne BSs and UAVs.
In particular, narrowband systems that provide position information or transmit control signals, such as the VHF Data Exchange System (VDES), may also be integrated for intelligent configuration of the heterogeneous MCN.
Additionally, it is essential to effectively allocate spectrum and power resources based on the characteristics of different service requirements \cite{p1004}--\cite{p1013}.
\textcolor{black}{Typical maritime communication services, such as geographic information services for safe navigation and video downloading services for passenger infotainment, have various requirements for bandwidth, latency and reliability. We will discuss the key technologies for service provisioning in detail in the following section.}

%Let us denote the beamforming matrix by $\mathbf{B}=[{{\mathbf{b}}_{1m}},{{\mathbf{b}}_{2m}},...,{{\mathbf{b}}_{Km}}]$, and the received signal vector at the $K$ users and the $k^{th}$ user can be expressed as
%\begin{align}
%\mathbf{y}={{\mathbf{H}}^{H}}\mathbf{x}+\mathbf{z}={{\mathbf{H}}^{H}}\mathbf{Bd}+\mathbf{z},
%\end{align}
%\begin{align}
%{{\mathbf{y}}_{k}}=\mathbf{H}_{k}^{H}{{\mathbf{B}}_{k}}{{\mathbf{d}}_{k}}+\left( \sum\limits_{k'=1,k'\ne k}^{K}{\mathbf{H}_{k}^{H}{{\mathbf{B}}_{k'}}{{\mathbf{d}}_{k'}}} \right)+{{\mathbf{z}}_{k}},
%\end{align}
%where $\mathbf{d}$ and ${{\mathbf{d}}_{k}}$ denote the data symbols, while $\mathbf{z}$ and ${{\mathbf{z}}_{k}}$ represent the additive white Gaussian noise.

%
%\begin{figure} [htb]
%\begin{center}
%\includegraphics*[width=9cm]{newhaiyu333.eps} %这行为导入图片文件名称
%\end{center}
%\vspace*{-4mm} \caption{Average downlink transmit power per user $P_{avg}$ versus the maximum transmit power $P_{max}$ for different user scheduling schemes.}\label{fig:2}  %这行为文章中显示图片的标题
%\vspace*{-2mm}
%\end{figure}

%\section{Main challenges of BMCNs}\label{sec:2}

\begin{figure*} [t]
\begin{center}
\includegraphics*[width=16cm]{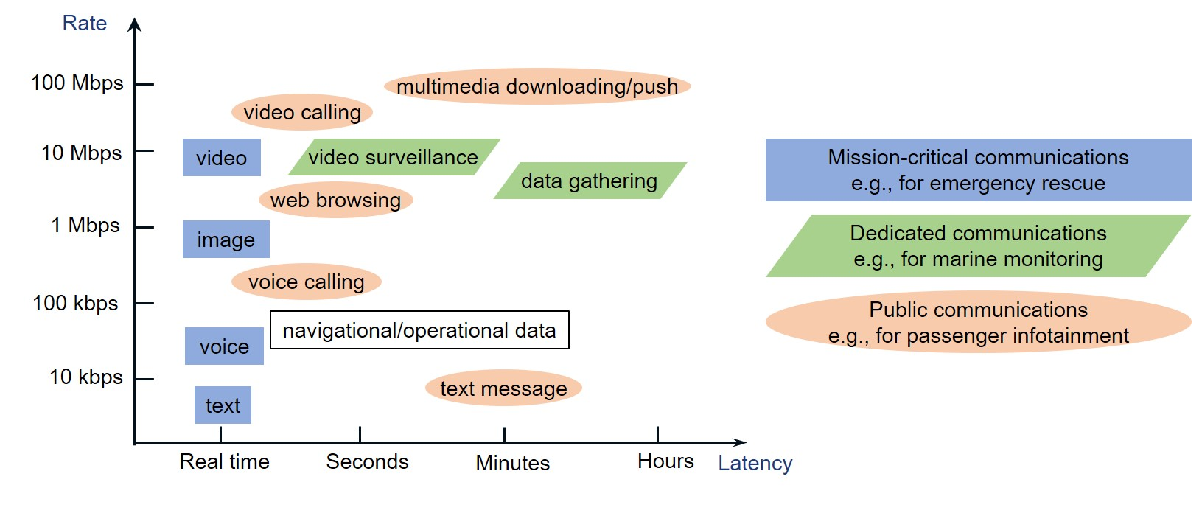} %这行为导入图片文件名称
\end{center}
\vspace*{-4mm}
\caption{Three types of maritime communication services and their required rate and latency.}\label{fig:2}  %这行为文章中显示图片的标题
%\vspace*{-2mm}
\end{figure*}

\section{Service Provisioning for Maritime Applications\label{sec:2}}

\subsection{Demand for Maritime Communications}

The demand for maritime communications emerged in the early 20th century.
Due to several maritime accidents, such as the sinking of the Titanic in 1912, the maritime community was awakened to the need for maritime communications in the event of search and rescue and to ensure the safety of ships and lives on the sea.
In 1914, the International Convention on the Safety of Life at Sea (SOLAS) was developed.
It mandates that ships sailing at sea must have battery-powered transceivers for transmitting and receiving radio alarm signals \cite{p5302}.
After that, maritime communications played an important role in distress communication and rescue.
%, and
%successfully completed numerous maritime rescue missions.
In this case, low-speed maritime communication services were enough to meet the demand for emergency rescue.

The number of maritime activities has increased dramatically since the beginning of the 21st century,
owing to the development of the world's economies and the prosperity of the modern shipping industry.
Maritime activities, such as marine tourism, offshore aquaculture and oceanic mineral exploration, have generated huge demand for high-speed and ultra-reliable maritime communication services. For example, the annual throughput of communication services provided by maritime satellites was less than 5 Gbps in 2005, while this number increased to approximately 66 Gbps in 2016 \cite{p1308}.
%The marine informatization construction involves industries such as maritime affairs, fisheries, ports, shipping, and coastal defense.
%Each maritime application has its unique and ever-increasing demand for maritime communication services.

If one takes a bird's-eye view over the ocean, one will find various types of marine users requiring communication services.
For sailing vessels, navigational and operational communication services are required for safe navigation.
For passengers, crew, fishermen and offshore workers, web browsing and multi-media downloading services are needed for their entertainment. The beacons are deployed to collect and upload meteorological and hydrological information, and platforms for oil exploitation require real-time operational data services. In particular, when a marine accident happens, real-time video communications
will be of great help for rescue.

In industrial applications, marine informatization management requires wireless data services, such as video surveillance, video conferencing, and navigational data services. Other marine industries, such as marine fisheries and offshore oil exploration, also have a large amount of data for uploading.
For marine tourism applications, multimedia services are needed to satisfy the passengers and the crew, and internet services are required to keep them connected at any time. For all of the applications described above, low-cost high-speed maritime communications are beneficial \cite{p1802}--\cite{p1803}.

%Data communications, including real-time data communications, require high-speed and low-cost maritime wireless communications.
%In terms of public communication applications, shipping is an important way of freight and passenger transportation. Other than receiving wireless signals from terrestrial communications operators in a few offshore areas, in most time, only satellite communications with high price and limited bandwidth are available. If the maritime wireless broadband network can be extended beyond the sea, it can not only provide services to the passengers and the crew in the leisure time, but also keep them connected to the Internet at any time \cite{p1803}.
%For example, residents living on isolated islands need telephone and television services, tourists on ships have demands for data services such as the Internet, and fishermen in the sea require more video services.

On the other hand, for maritime rescue, real-time and high-reliability maritime communication services are required to enable coordination between ships and between ship and shore. In addition to text and voice, real-time video communications will be very helpful for conducting rescue operations in a more accurate manner.
Real-time and high-reliability services are also required for maritime military applications, e.g., for communication and coordination between warships and between fleets and land.
A higher level of security is also required in these applications to prevent the data transmission from being intercepted by eavesdroppers \cite{p1804}.

Based on the nature of the communication network organization and service demands, maritime communication services can be classified as secure communications, dedicated communications and public communications \cite{p202}. Secure communications include voyage reports and severe weather warnings to ensure safe navigation, as well as communications for help, search and rescue in the event of a shipwreck.
%According to the criticality of the situation, in the communication, the distress (calling for help) signal (telephone is "SOS", the phone is "MAYDAY"), the emergency signal (telephone is "XXX", the phone is "PAN PAN"), safety signal ( The telegram is "TTT" and the phone number is "SECURITE").
%Communication with a distress signal is also known as “distress communication” and enjoys the highest priority. Other communications must not interfere. Coast stations specializing in secure communications are usually maritime security agencies in countries and do not charge for communication.
Dedicated communications allow the navigation department or the maritime enterprise to establish internal communication protocols, and set up communication links between a self-designed or leased coast station and its own ships according to the application requirements.
Public communications refer to the communications between ship personnel, passengers, and any users of the land-based public communication network \cite{p1501}.
%Coast stations that provide public communications services are usually operated by national telecommunications departments or enterprises, charging communication fees, and assuming the obligation of non-charged secure communication services .

Typical maritime communication services are depicted in Figure 8, according to their requirements for rate and latency \cite{p222}.
%As is depicted,
Navigational and operational communication services, such as ship reporting, voyage reporting, electronic navigational chart (ENC) updates, coast state notification, and environment notification, are required by all vessels.
These services do not require large bandwidth and can be provided by coast stations or maritime satellites \cite{p713}\cite{p716}.
Secure communications services, such as those for emergency rescue and military missions, have a critical demand for latency.
There have been increasing demands for real-time video communications in addition to voice services in such mission-critical maritime activities \cite{p703}--\cite{p711}.
Dedicated communications usually require a large bandwidth but tolerate high latency.
%The demands for dedicated services keep increasing with the development of maritime industries, such as marine monitoring and maritime oil exploitation.
Public communications, such as web browsing and video downloading services, are mainly for passenger and crew infotainment.
For public communications, a great deal of bandwidth is required, while the demand for latency varies from real-time to minutes.
%The demands for maritime public communications have grown explosively, and will continue to increase with the development of maritime tourism, as well as smartphones.
%%Typical maritime communication services and the requirement of rate, latency and criticality are depicted in the table below \cite{p221}.
%%Different services require different communication rates and signal integrity, while terrestrial cellular networks tend to make a compromise.
%From the above, the maritime application scenarios are quite different, and the service requirements are also unique.
%%Therefore,
%To provide such diversified services of high rate, low latency and high reliability for different applications within one network is a major challenge for MCNs.
%%How to effectively allocate spectrum and power resources to improve the spectral and energy efficiency of the network based on the characteristics of different users' service requirements is also a problem that needs further study.
%In this case, effective resource allocation and efficiency in spectrum and energy are very important.

%ENC: Electronic Navigational Chart

%VTS: vessel traffic system

%\begin{figure} [htb]
%\begin{center}
%\includegraphics*[width=10cm]{service.eps} %这行为导入图片文件名称
%\end{center}
%\vspace*{-4mm} \caption{Maritime application requirements.}\label{fig:2}  %这行为文章中显示图片的标题
%\vspace*{-2mm}
%\end{figure}

\subsection{Service Provisioning for Typical Maritime Applications}

Different services have different requirements for bandwidth, latency and reliability, as depicted in Figure 9.
For example, the data services for maritime rescue and operation of oil platforms have a critical demand for real-time video and high reliability. The multimedia downloading and data gathering services usually require a large bandwidth but tolerate high latency.
One unique property of maritime services is that
the density of vessels is low, while the users (passengers/crew/fishermen/offshore workers) are clustered in a small area (ship/platform).
Therefore, the passenger/crew infotainment services are sparsely distributed on the sea, but densely clustered in each vessel.

In view of this, next we will introduce the existing and potential schemes to provide maritime-specific services. In addition to the intelligent navigational communication services required by all vessels, the clustered distribution services, delay/reliability-sensitive maritime services, as well as delay-tolerant maritime services will be discussed.

\subsubsection{Intelligent Navigational Communication Services}

To achieve safe navigation and improve shipping efficiency, sailing ships need to be provided with real-time and accurate maritime traffic information in the fastest and most efficient way.
In \cite{p701}, the authors considered the feasibility of utilizing illuminators sent by Inmarsat for maritime surveillance and navigation, especially for marine obstacle avoidance. In \cite{p717}, the authors introduced the utilization of satellite-based AIS receivers to extend traffic monitoring zones to open seas, as well as collision avoidance in high traffic zones.
More recently, the authors of \cite{p702} proposed a parallel signal processing architecture and algorithms for satellite-based AIS to cope with the message collision in dense maritime zones and reduce the downlink power, bandwidth, and latency.
\textcolor{black}{Similar to satellite-based AIS, the Long-Range Identification and Tracking (LRIT) system is a real-time reporting system that allows for ship detection and identification from space \cite{p20201129-0}.}

\textcolor{black}{
In addition to using the above-mentioned satellite-based systems, navigational communication services can also be provided by shore-based systems, such as terrestrial AIS and coastal radars \cite{p20201129-1}. Shore-based radar systems can promote safe navigation by collision monitoring and grounding prevention. Particularly, the authors of \cite{p20201129-2} reported the experimental performance assessment of high frequency surface wave (HFSW) radars, which have wider
coverage than conventional microwave radars. The detection capabilities
of HFSW radars were evaluated and enhanced using spectrum analysis techniques in \cite{p20201129-3}.}
In short, an intelligent maritime transportation network is currently formed by making full use of the Geographic Information System (GIS), the Global Positioning System (GPS), remote sensing (RS), and other technologies \cite{p1506}--\cite{p1508}.

\subsubsection{Passenger/Crew Infotainment--Clustered Distribution Services}

Different from the terrestrial scenario, where users are scattered on the land,
in the MCN, the density of vessels is low, while the users (passengers/crew/fishermen/offshore workers) are clustered in a small area (ship/platform) \cite{p10282}.
Therefore, the passenger/crew infotainment services are sparsely distributed on the sea, but densely clustered in each vessel.
To provide such services, phased array antennas and beamforming techniques can be used. The direction
of beams can be controlled according to the location of vessels, which can be obtained from the AIS \cite{p1020}.
In \cite{p1013}, the authors proposed a user-centric communication structure and an antenna selection scheme based on distributed  antennas. In \cite{p1701}, the authors proposed a location-aware dynamic beam scheduling scheme to provide users in each ship with guaranteed QoS to strike a balance between the throughput and fairness among different ships.

% Table generated by Excel2LaTeX from sheet 'service'
\begin{table*}[htbp]
  \centering
  \caption{Technologies for Guaranteeing Typical Maritime Communication Services.}
  \scriptsize
    \begin{tabular}{|p{8.24em}|p{8.675em}|l|l|}
    \hline
    \multicolumn{1}{|l|}{\textbf{Service}} & \multicolumn{1}{l|}{\textbf{Characteristics}} & \textbf{Solutions} & \textbf{Technologies} \bigstrut\\
    \hline
    \multirow{7}[6]{*}{\shortstack{intelligent navigation \\ and emergency rescue}} & \multirow{7}[6]{*}{\shortstack{real-time, \\ high reliability,\\ high accuracy}} & \multicolumn{1}{l|}{\multirow{4}[6]{*}{\shortstack{using satellite-based \\positioning and geographic \\information systems}}} & utilizing illuminating signals sent by Inmarsat for maritime surveillance and navigation \cite{p701} \bigstrut\\
\cline{4-4}    \multicolumn{1}{|l|}{} & \multicolumn{1}{l|}{} &       & utilizing satellite-based AIS receivers to extend traffic monitoring zones \cite{p717} \bigstrut\\
\cline{4-4}    \multicolumn{1}{|l|}{} & \multicolumn{1}{l|}{} &       & parallel signal processing for satellite-based AIS to cope with message collision \cite{p702} \bigstrut\\
\cline{4-4}    \multicolumn{1}{|l|}{} & \multicolumn{1}{l|}{} &       & \textcolor{black}{collecting and disseminating vessel position information via the self-reporting-based LRIT system \cite{p20201129-0}} \bigstrut\\
\cline{3-4}    \multicolumn{1}{|l|}{} & \multicolumn{1}{l|}{} &  \multicolumn{1}{l|}{\multirow{3}[6]{*}{\textcolor{black}{\shortstack{using shore-based \\positioning and geographic \\information systems}}}}     & \textcolor{black}{collision monitoring and grounding prevention using terrestrial AIS and coastal radar systems \cite{p20201129-1}} \bigstrut\\
\cline{4-4}    \multicolumn{1}{|l|}{} & \multicolumn{1}{l|}{} &       & \textcolor{black}{experimental performance assessment of HFSW radars \cite{p20201129-2}} \bigstrut\\
\cline{4-4}    \multicolumn{1}{|l|}{} & \multicolumn{1}{l|}{} &       & \textcolor{black}{enhancing detection capabilities of HFSW radars using spectrum analysis techniques \cite{p20201129-3}} \bigstrut\\
    \hline
    \multirow{3}[6]{*}{\shortstack{passenger and crew \\infotainment}} & \multirow{3}[6]{*}{\shortstack{low density of vessels,\\ users clustered \\in a small area}} & \multicolumn{1}{l|}{\multirow{3}[6]{*}{\shortstack{generating directional \\beams according to \\the location of vessels}}} & fairness-oriented beam scheduling using phased-array antennas \cite{p1701} \bigstrut\\
\cline{4-4}    \multicolumn{1}{|l|}{} & \multicolumn{1}{l|}{} &       & \multicolumn{1}{p{39.315em}|}{antenna selection based on distributed antennas \cite{p1013}} \bigstrut\\
\cline{4-4}    \multicolumn{1}{|l|}{} & \multicolumn{1}{l|}{} &       & \multicolumn{1}{p{39.315em}|}{antenna tracking \cite{p1020}} \bigstrut\\
    \hline
    \multirow{3}[6]{*}{\shortstack{multimedia \\downloading and \\data gathering}} & \multirow{3}[6]{*}{\shortstack{large data volume,\\high latency tolerance}} & \multicolumn{1}{l|}{\multirow{3}[6]{*}{\shortstack{service scheduling based \\on the mobility of vessels}}} & offline scheduling based on the deterministic network topology \cite{p503} \bigstrut\\
\cline{4-4}    \multicolumn{1}{|l|}{} & \multicolumn{1}{l|}{} &       & opportunistic routing based on lane intersecting opportunities \cite{p502} \bigstrut\\
\cline{4-4}    \multicolumn{1}{|l|}{} & \multicolumn{1}{l|}{} &       & store-carry-and-forward scheduling \cite{p1022} \bigstrut\\
    \hline
%    \multirow{4}[8]{*}{\shortstack{underwater \\communications}} & \multirow{2}[4]{*}{\shortstack{high bandwidth,\\low latency}} & optical communications & Aquatec AQUA modem [249], MIT low power led modem [250], Sonardyne BlueComm 500 [251] \bigstrut\\
%\cline{3-4}    \multicolumn{1}{|l|}{} & \multicolumn{1}{l|}{} & EM communications & CoSa WiFi [252], INESC TEC Dipole [253], WFS Seatooth S500 [254] \bigstrut\\
%\cline{2-4}    \multicolumn{1}{|l|}{} & \multicolumn{1}{l|}{long range} & acoustic communications & Arctic transmission [255], BaltRobotics Prototype [256], LinkQuest UWM series [257] \bigstrut\\
%\cline{2-4}    \multicolumn{1}{|l|}{} & \multicolumn{1}{l|}{high security} & quantum communications & in trials \bigstrut\\
%    \hline
    %\shortstack{maritime IoT \\applications} & \shortstack{long range, \\massive connection } & designing LPWANs & LoRa and SigFox (using unlicensed spectrum), NB-IoT (working on licensed spectrum) [248] \bigstrut\\
%    \hline
    \end{tabular}%
  \label{tab:addlabel}%
\end{table*}%

\subsubsection{Mission-critical Services with Low Latency and High Reliability}

When a marine accident happens, real-time and high-reliability communication services are important for maritime search and rescue.
The emergency communication systems based on UAVs and low-orbit satellites can provide real-time transmissions of voice, image, video, etc., and help improve the communication security in the remote area.
%and  Efficient search and rescue command work provides a strong guarantee.
In addition, underwater emergency communications can also provide communication and location services for underwater rescue, wreck positioning, as well as search and salvage \cite{p6009}.

For ship-to-ship communications between the rescue team and the ship with accident, it is worth mentioning that the height and angle of the ship-borne antenna are rapidly changing with the fluctuation of the sea surface. Thus, the maritime channel fading is particularly sensitive to antenna height and angle, which may cause frequent link interruption. To cope with this challenge, antenna switching techniques have been proposed in \cite{p6008}. In this paper, when the rocking angle of a ship is more than a threshold, antenna switching will be triggered to improve link stability and packet delivery ratio.

%In maritime activities such as marine monitoring and maritime defense as well as some special operations, ensuring the security of communications and preventing information from being intercepted is particularly critical. It is worth noting that due to the limitation of the BS site of the maritime communication network, in order to implement wide area coverage with a limited number of BSs, the transmission power of the BS usually needs to be set high, which makes the data transmission more easily intercepted by eavesdroppers. It has brought great challenges to the design of the security of maritime communications networks.

%Physical layer security technology ensures information security by utilizing different physical characteristics of different wireless channels, and is considered as an alternative to traditional encryption technology \cite{p801}\cite{p802}. Currently, the physical layer security technology has been widely used in various network topologies and eavesdropping scenarios, such as multiple antennas \cite{p803}-\cite{p805}, multiple users \cite{p806}-\cite{p808}, relays \cite{p809}, and cognitive radio networks \cite{p810}-\cite{p812}.
%However, in the maritime communication network, the analysis and design of the physical layer security will be completely different from the traditional terrestrial communication network due to the differences in channel models, network architecture, and user behavior.

\subsubsection{Multimedia Downloading and Data Gathering--Delay-tolerant Services}

The downloading of multi-media and the uploading of hydro-meteorological information require broadband communications and high latency tolerance \cite{p1513}. To provide this kind of delay-tolerant services, several resource scheduling methods for maritime communications have been proposed.
%The existing resource scheduling methods for maritime communications are mainly upper layer algorithms for video, weather/hydrological distribution and other delay-tolerant services, and there is still no physical layer resource scheduling algorithm [24][25].
In \cite{p502}, an opportunistic routing scheme for delay-tolerant MCNs based on lane intersecting opportunities was proposed.
In \cite{p503}, three offline scheduling algorithms for video uploading in MCNs based on the deterministic network topology were proposed.
These studies utilized the predictability and stability of marine user movement \cite{p1022}.
%Therefore,
%For future work, it is promising to explore the mobile law of offshore users, expand the scope of resource scheduling in the time domain, and explore resource scheduling frameworks and algorithms for wide-area mobile coverage.
%To this end,

%A new framework for joint optimization of service services and BS transmission can be built. For example, a proxy can be set up for each user, and the agent collects and combines information such as the user location, shipping route, service demand, and link resource status of the BS, along with the sea status. The resource scheduling is performed by estimating user location and CSI, and then these data are sent to the BSs and to the users.

In Table VII, we summarize the technologies used to provide typical maritime communication services.
The above work has designed different networks for different maritime service requirements. However, for maritime communications, due to the limited number of BSs, the best solution is to establish a network that supports all types of services at sea. To this end, network resources should be flexibly coordinated to present different performance gains according to different service requirements. Therefore, cross-layer design and joint optimization of the physical layer, MAC layer, and network layer are required for multiple service types and QoS requirements, such as different rates, delays, and reliability, with comprehensive consideration of channel status and service requirements.
\textcolor{black}{
\subsection{Low Power Communications for Maritime IoT}
}
\textcolor{black}{
The Low-Power Wide-Area Network (LPWAN) is designed for low-bandwidth, low-power, long-range, and massively connected IoT applications, which can support data collection from sensors up to several tens of kilometres from shore \cite{p20201206-1}.
LPWAN has wider coverage than other wireless connection technologies (such as Bluetooth and Wi-Fi) and lower power consumption than cellular technologies (such as 4G/5G).
LPWAN can be divided into two categories: one using unlicensed spectrum, such as LoRa and SigFox, and
the other using licensed spectrum, such as NB-IoT.
%LoRa uses the 433/868/915 MHz frequency band, providing transmission rates up to several hundred kbps with a coverage of 15 km.
%SigFox uses the 868 MHz band and the 915 MHz in Europe and America and has a coverage of 3--10 km.
NB-IoT has higher-reliability due to less interference in the licensed band, while its power consumption is higher than the other two technologies with a coverage of approximately 10 km \cite{p10511}. The role of LPWAN in cellular IoT to support massive machine-type communications (mMTC) is being discussed for beyond 5G networks, while it remains to be seen whether its coverage can be further expanded for maritime IoT \cite{p20201213-6}.}
\\

\begin{figure*} [htb]
\begin{center}
\includegraphics*[width=14cm]{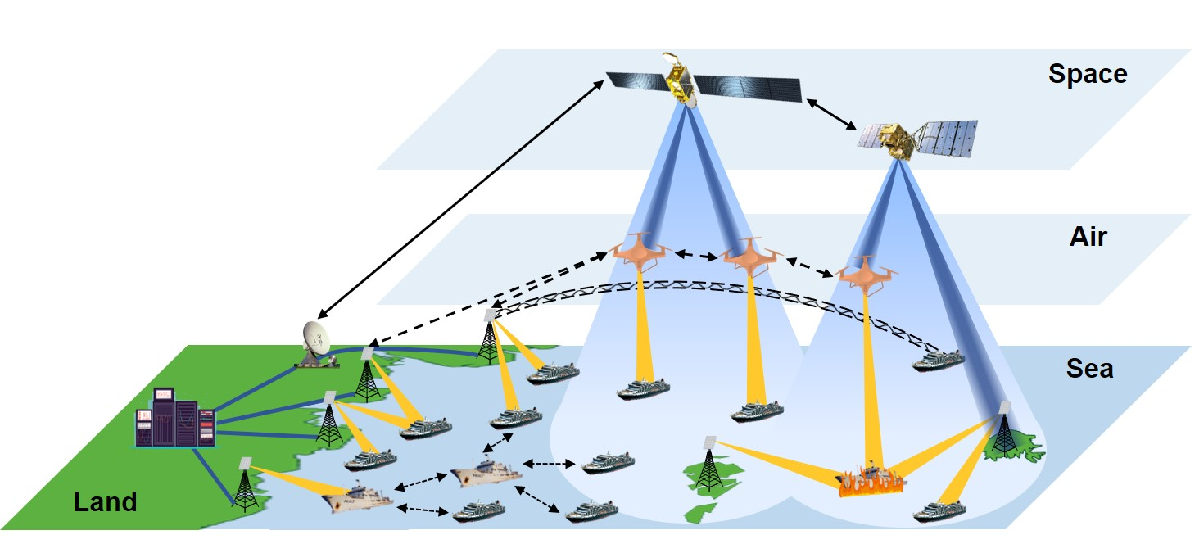} %这行为导入图片文件名称
\end{center}
%\vspace*{-4mm}
\caption{Future MCNs: satellite-air-ground integrated, environment-aware and service-driven.
}\label{fig:2}  %这行为文章中显示图片的标题
%\vspace*{-2mm}
\end{figure*}

\subsection{Cross-Layer Design for QoS-guaranteed Maritime Communications}

It should be noted that the maritime application scenarios are quite different and that the marine service requirements are unique.
To adapt to the different types of services and different requirements for QoS, it may be necessary to jointly consider the CSI and service requirements by designing cross-layer optimization schemes \cite{p10411}.
For example, potential synergies of exchanging information between different layers for real-time video streaming in ad hoc networks was explored in \cite{p10412}. The joint design of physical, MAC, and network layers was considered for interference-limited wireless sensor networks in \cite{p10413}.
Reference \cite{p10414} analysed the cross-layer design of QoS-forward geographic wireless sensor network routing strategies in green IoT.
%Currently, the fixed full coverage scheme is difficult to support the service requirements of broadband services.
%In order to overcome the above problems,
In particular, service-driven methods can be used in MCNs to allocate resources, and user-centric transmissions can be used to implement rapid link-building services \cite{p1509}. Resource conditions and service requirements can be exploited for flexible resource allocation for different services  \cite{p1503}.
 %are provided , and communication services are efficiently provided for the service quality requirements of different services.
%Possible technical approaches.
\textcolor{black}{In addition, a programmable architecture based on SDN is believed to be useful for the maritime IoT \cite{p20201206-5}.}

Using these service-driven schemes, it is possible to comprehensively address the dynamic changes in the location and demand of marine users, as well as the wide range of maritime network coverage with severely limited resources. Moreover, a new framework for joint optimization of service scheduling and BS transmission can be built. For example, a proxy can be set up for each user, and the agent collects and combines information such as the user location, shipping route, service demand, and link resource status of the BS, along with the sea status. The resource scheduling is performed by estimating user location and CSI, and then these data are sent to the BSs and to the users.

\begin{figure*} [htb]
\begin{center}
\includegraphics*[width=13cm]{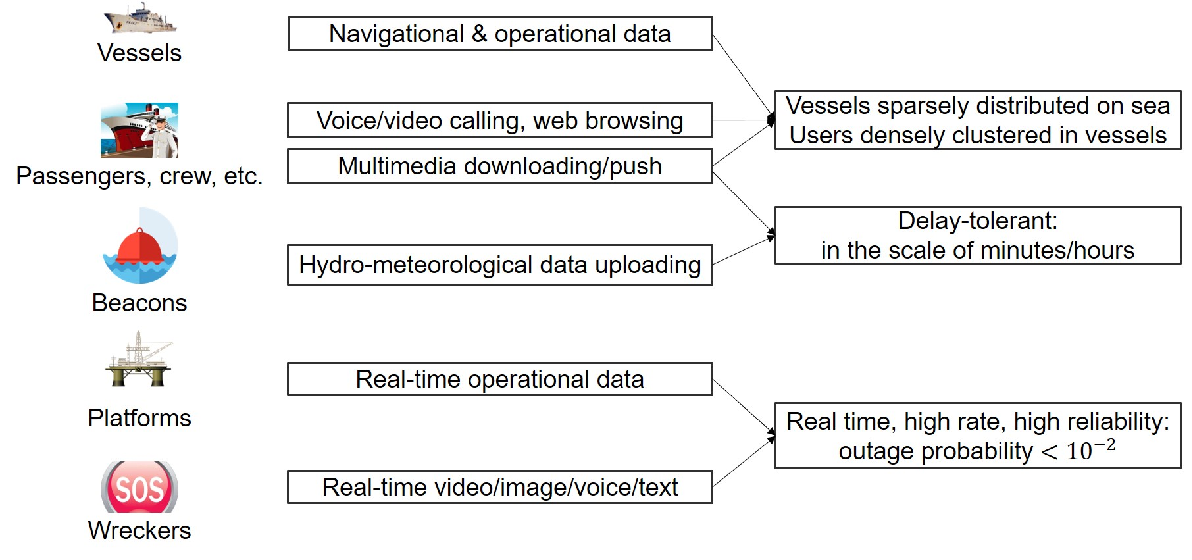} %这行为导入图片文件名称
\end{center}
\vspace*{-1mm}
\caption{Maritime-specific communication services and their features.}\label{fig:2}  %这行为文章中显示图片的标题
%\vspace*{-2mm}
\end{figure*}

\section{Architecture and Features of Future MCNs}

\textcolor{black}{In the last three sections, we have discussed the key technologies for enhancing transmission efficiency, increasing broadband coverage, and providing domain-specific services for the MCN.
As discussed in Section IV,
%in addition to maritime satellites, shore \& island-based BSs, UAVs, high-end ships and offshore lighthouses can be exploited to serve as maritime BSs.
%In order to give full play to the advantages of the above-mentioned means of maritime coverage and overcome their shortcomings,
%overcome the contradiction between the breadth and strength of the coverage of the broadband maritime communication network,
%it is a feasible approach that
a heterogeneous network is useful which requires the coordination of terrestrial and non-terrestrial BSs \cite{p1511}.} In this network, the terrestrial BSs mainly cover the offshore waters, and the satellites mainly cover the ocean areas. At the same time, the ship-borne BSs on the sea can be used as relay nodes to serve nearby vessels.
To facilitate the use of the abovementioned BSs, advanced hardware needs to be developed, such as new antennas with higher directivity and lower complexity, radio frequency (RF) amplifiers with higher linearity and lower noise, as well as airborne and shipborne equipment with lighter weight and lower power consumption \cite{p6001}\cite{p6003}.
Future MCNs should enhance the transmission efficiency in the complex and varied maritime environment, extend the coverage by taking advantage of and overcoming the shortcomings of different coverage methods, and develop service-specific transmission and coverage techniques to meet the unique service requirements of marine users.

\subsection{Requirements and Characteristics of Future MCNs}

To improve transmission efficiency, future MCNs need to be aware of the environment, such as the sea level, temperature, humidity, and wind speed, and use this awareness to obtain more accurate prediction of the CSI and adopt more efficient transmission techniques that counteract the dynamic changes in maritime channels \cite{p1514}.
In addition, future MCNs need to be able to provide flexible services based on resource conditions and service requirements.
This can address the dynamic changes in the location and demand of marine users and thus allow the provision of dynamic and on-demand coverage using limited resources.

As depicted in Figure 10, future MCNs can adopt more flexible coverage modes and service patterns by utilizing the unique maritime channels and service characteristics.
Specifically, the environmental information, positional information and service information can be collected by narrowband systems and exploited by the central processor (and BSs serving as edge processors \cite{p6004}\cite{p6005}) to design integrated satellite-air-ground systems.
%For example, a signaling subnet is constructed by using the wide-area coverage capability of the VHF or the narrowband satellites to collect the service information and the space-time location of users. At the same time, based on information such as the distribution characteristics of the marine users, service requirements, and CSI, the service capabilities will be improved by dynamically allocating resources such as spectrum, power, and beams.
For example, a long-distance communication link can be dynamically established, depending on whether the user is in an environment that satisfies the conditions under which the evaporating duct exists.
When high-speed and high-reliability communication services are required for rescuing a vessel on fire, the nearby vessels and UAVs can gather together to provide ship-borne and air-borne services.

\begin{figure*} [t]
\vspace*{1mm}
\begin{center}
\includegraphics*[width=14.5cm]{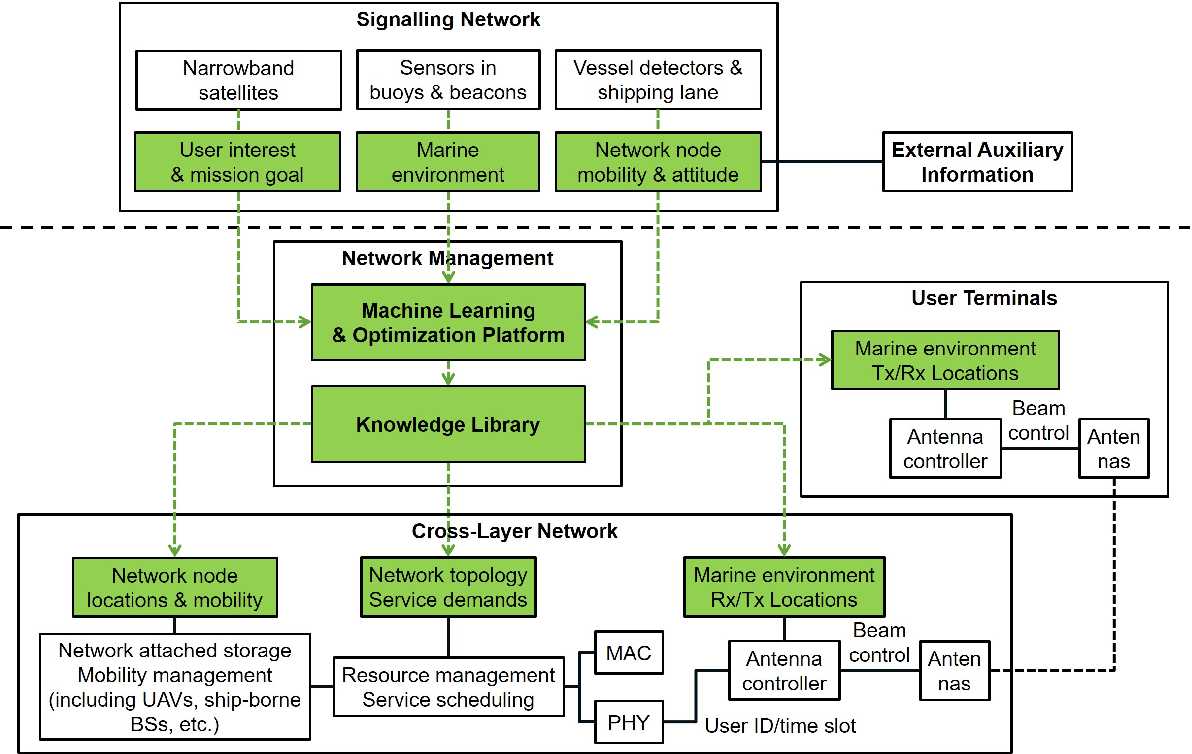} %这行为导入图片文件名称
\end{center}
\vspace*{-1mm}
\caption{Exploiting the knowledge library for intelligent coverage and transmission in the future MCN.}\label{fig:2}  %这行为文章中显示图片的标题
%\vspace*{-2mm}
\end{figure*}

\subsection{Exploiting the Knowledge Library for Intelligent MCNs}

Following the discussion in Section VII.A, it is recommended to establish a knowledge library
that contains all environmental information
for future MCNs. The knowledge library is used to portray the complex signal propagation environment, network topology, and service characteristics, based on which the transmission efficiency and coverage performance can be improved through optimization.
The knowledge library comes from both internal information on the communication process, such as the CSI, and external information, such as the maritime environment, network node position, and user behaviour characteristics, as depicted in Figure 11.
The external information can be gathered by buoys, ship-borne sensors, etc., and then uploaded to the central processor via narrowband systems. In the machine learning and optimization platform with the central processor, machine learning techniques can be adopted to jointly process the internal and external information and establish the knowledge library, including the hierarchical maritime channel model, the network topology evolution model, and the service model. The available BSs with extra storage capacity and computing power can be exploited to serve as edge processors. \textcolor{black}{Neural network structures \cite{p20201213-5} and federated learning technologies \cite{p20201213-4} being discussed for the sixth-generation (6G) network may contribute to the above process.}

Utilizing the knowledge library, the machine learning and optimization platform will further perform transmission optimization, network management, and service scheduling for the MCN.
For example, in transmission optimization, using the meteorological and hydrological information gathered by maritime buoys and weather satellites, the machine learning and optimization platform can model the temporal-spatial distribution of maritime channels and add it to the knowledge library.
Then, the MCN can predict the existence of deep-fading channels due to the 2-ray/3-ray propagation characteristic in maritime scenarios and overcome the deep fading using diversity techniques.
The MCN can also estimate when and where an evaporation duct will exist and dynamically configure the network and allocate resources for more efficient transmissions.

In network management, the BSs are often static in traditional schemes. Hence, they fail to adapt to the dynamic network topology in the MCN.
Using the knowledge of network node mobility, such as the shipping lane information obtained from the AIS, and the attitudes of satellites and UAVs, the machine learning and optimization platform can construct a network topology evolution model from BS/user position prediction. Based on that, the network can be dynamically and intelligently configured for wider coverage, as represented by the irregularly configured heterogeneous network in Figure 10.

In service scheduling, based on user interest and mission goals, the machine learning and optimization platform can establish a personalized service model, characterizing the distribution of service occurrence time, the length of service duration, and the service requirement. Using the knowledge library, the network can perform service forecasting, provide user-specific services, and dynamically adjust the allocation of resources in the case of emergency. In general, statistical service models can be applied for resource allocation to greatly improve the service capabilities of the MCN.
Figure 11 shows all the scenarios discussed.

\subsection{Open Problems}

Based on the lessons from Section III-V, to facilitate the construction of future MCNs supporting more intelligent coverage and transmission, as well as providing higher QoS,
it is important to address the environment-sensitive maritime channels, to make coordination of all available coverage methods, and to adapt to the service
demands from maritime applications.
In particular, the external auxiliary information can be exploited to establish an intelligent MCN to achieve on-demand agile coverage and efficient transmissions. On the other hand, this new framework also poses challenges for both communications theories and practices. We list some open issues as follows.

First, the new maritime channel model is essential for transmission efficiency enhancement. In most traditional applications, we deal with wireless channels from a mathematical perspective, e.g., we treat the channel coefficient as a random variable and use lots of measurements to derive a statistical channel model. For the intelligent MCN framework, we need to treat the wireless channel from a more physical perspective. An environment-sensitive maritime channel considering sea surface conditions and atmospheric conditions is required \cite{p10501}. Towards this end, a new channel measuring method is needed, which has to synchronize the channel measurement with meteorological observation. Based on that, the physical model of the meteorological information can be integrated into mathematical statistical analysis to obtain an external auxiliary information-driven maritime channel model. There exists hierarchy in the new channel model, and accordingly, structural modelling is a potential solution to integrate both the physical and the mathematical features. Structural processing is also a potential solution to realize environment-aware transmission enhancement by matching the hierarchy of the channel.

Second, the coordination of all available wireless techniques is another important open issue.
\textcolor{black}{It is desired to integrate satellites and high-altitude platforms into 6G to expand its coverage, which may facilitate the development of MCNs \cite{p20201213-1}.}
In general, satellites, UAVs and terrestrial BSs are quite different in terms of both mobility and transmission performance. The satellite follows the dynamics of orbits, leading to relatively static GEOs and fast moving LEOs. The movement of the UAV is more agile than satellites. However, its mobility may be significantly restricted by weather conditions over the sea. When the terrestrial BS is equipped on vessels, it can move, but its mobility is determined by the fixed shipping lane of the corresponding vessel. All these issues make the stochastic topology of MCNs a challenge \cite{p90001}\cite{add1}. Moreover, the transmission delay and rate are also quite different for satellites, UAVs and terrestrial BSs. This further complicates the network control of MCN.
To solve this problem, new coverage metrics can first be established theoretically. Different from traditional cellular networks, the coverage performance cannot be calculated by the sum of the achievable rate in different cells because the cellular structure may not hold in the maritime scenario. The coverage metric of a non-cellular MCN can also consider the distribution of users from a practical perspective. As maritime users are sparsely distributed on the ocean, it is not efficient to cover the whole geographical area as the cellular architecture does. A user distribution-aware coverage performance metric is desired, which needs both theoretical and practical research.

Third, new marine services also pose challenges in the design of intelligent MCNs. Due to the limited BS sites, the optimal solution is to build one network supporting all services over the sea. This is quite different from the terrestrial network, where we have already had a number of different networks for different services, e.g., the 4G/5G cellular network, the Wi-Fi network, as well as a variety of private networks. To achieve this, the intelligent MCN needs to be flexible enough to support both existing and upcoming communication services. Taking the fast-developing Maritime Autonomous Surface Ships (MASS) as an example, it requires both high-speed multimedia services for video surveillance and ultra-reliable services for remote control. This is challenging when using only one system.
\textcolor{black}{The 5G cellular network could support three usage scenarios with different service types, including enhanced mobile broadband (eMBB), ultra-reliable low-latency communication (URLLC), and mMTC \cite{p20201213-2}.
In the future, the network resource ought to be orchestrated in a more flexible and agile manner, thus presenting different performance gains according to different requirements. Both mobile edge computing \cite{add0725} and block-chain technologies \cite{p90002}\cite{p20201213-3} being discussed for 6G can be used in this orchestration, although it remains open.}

%To facilitate the construction of future MCNs supporting more intelligent coverage and transmission, as well as providing higher QoS,
%it is important to address the environment-sensitive maritime channels, to make coordination of all available coverage methods, and to adapt to the service
%demands from maritime applications.
%Some open problems are listed below from the above three aspects:
%\begin{itemize}
%  \item Precise measurements and structural modelling for maritime channels, considering sea surface conditions and atmospheric conditions
%  \item Smart channel estimation, environment-aware transmission and resource management technologies for MCNs
%  \item Architecture, optimization models, and design criteria of hybrid satellite-terrestrial MCNs
%  \item Coverage performance analysis and enhancing technologies for MCNs, such as Internet of Vessels (IoV) and UAVs-enabled maritime coverage
%  %\item Routing methods and protocols for maritime mesh/ad hoc networks
%  %\item Interference analysis, alignment, avoidance, and coordination in MCNs, including cognitive radio technologies
%  \item Cross-layer design for providing guaranteed QoS for marine users, such as Maritime Autonomous Surface Ships (MASS), and physical layer security issues for mission-critical services
%  \item Artificial intelligence methods for smart MCNs, such as machine learning, mobile edge computing, and block-chain technologies
%\end{itemize}
%}

\section{Conclusions }\label{sec:4}

This paper has provided a comprehensive review of hybrid satellite-terrestrial MCNs for the maritime IoT, including the demand for maritime communications, state-of-the-art MCNs, and enabling technologies.
It has been recognized that a large performance loss is usually inevitable if the existing 4G/5G and satellite communication technologies are used directly for the maritime scenario.
Thus, conventional communication theories and methods need to be tailored to match the unique characteristics of MCNs in terms of dynamic electromagnetic propagation environments, geometrically limited available BS sites and rigorous service demands from mission-critical applications.
Towards this end, we have categorized the enabling technologies into three types, i.e., enhancing transmission efficiency, extending network coverage, and provisioning maritime-specific services.
We have illustrated and compared the technologies in terms of their objectives, methods, and characteristics of MCNs used.
Facing the future, more research on communication and networking theories is still needed to avoid simple integration of existing networks.
We have accordingly envisioned the use of external auxiliary information to build up an environment-aware, service-driven, and integrated satellite-air-ground MCN. The corresponding open issues have also been discussed.

\end{document}